\documentclass[preprint,12pt]{elsarticle}

\usepackage{subcaption}

\usepackage[table]{xcolor}
\usepackage{graphicx}
\usepackage{amsthm}
 \usepackage{amsfonts}
\usepackage{rotating}
\theoremstyle{definition}
\usepackage{amssymb}
\usepackage{booktabs}
\usepackage{multirow}
\usepackage[hyphens]{url}
\usepackage{balance}

\journal{Journal of Information Security and Applications}
\begin{document}
\begin{frontmatter}
\title{Exploiting Statistical and Structural Features for the Detection of Domain Generation Algorithms}

\author[unipi,arc]{Constantinos Patsakis\corref{cor1}}
\ead{kpatsak@unipi.gr}
\author[unipi,arc]{Fran Casino}
\ead{francasino@unipi.gr}
\address[unipi]{Department of Informatics, University Piraeus, 80 Karaoli \& Dimitriou str, 18534 Piraeus, Greece}
\address[arc]{Information Management Systems Institute, Athena Research Center, Artemidos 6, Marousi 15125, Greece}

\begin{abstract}

Nowadays, malware campaigns have reached a high level of sophistication, thanks to the use of cryptography and covert communication channels over traditional protocols and services. In this regard, a typical approach to evade botnet identification and takedown mechanisms is the use of domain fluxing through the use of Domain Generation Algorithms (DGAs). These algorithms produce an overwhelming amount of domain names that the infected device tries to communicate with to find the Command and Control server, yet only a small fragment of them is actually registered. Due to the high number of domain names, the blacklisting approach is rendered useless. Therefore, the botmaster may pivot the control dynamically and hinder botnet detection mechanisms. To counter this problem, many security mechanisms result in solutions that try to identify domains from a DGA based on the randomness of their name.

In this work, we explore hard to detect families of DGAs, as they are constructed to bypass these mechanisms. More precisely, they are based on the use of dictionaries or adversarial approaches so the generated domains seem to be user-generated. Therefore, the corresponding generated domains pass many filters that look for, e.g. high entropy strings or n-grams. To address this challenge, we propose an accurate and efficient probabilistic approach to detect them. We test and validate the proposed solution through extensive experiments with a sound dataset containing all the wordlist-based DGA families that exhibit this behaviour, as well as several adversarial DGAs, and compare it with other state-of-the-art methods, practically showing the efficacy and prevalence of our proposal.
\end{abstract}


\end{frontmatter}
\section{Introduction}
The ceaseless efforts of malware authors to enhance cybercrime with sophisticated techniques \cite{MANSFIELDDEVINE201815} is creating a new ``business'' paradigm. Such a business has a myriad of monetisation sources \cite{isma2018,patsakis2020unravelling} including, but not limited to, ad injection \cite{CHEN2017164}, spamming \cite{rao2012economics}, denial of service \cite{sabillon2016cybercrime}, ransomware-based extorsion \cite{al2018ransomware} or phishing \cite{chiew2018survey}. In this regard, one of the most critical aspects of such malware campaigns is the control and management of the compromised hosts. This enables the malware author, apart from compromising the victim's security and privacy, to orchestrate further attacks and prolong the discovery of the attack.

In the past, the prevalent methodology was to establish a direct communication channel between the {\em Command and Control} (C\&C) server and the infected devices. However, this strategy had several flaws, since blacklisting a specific IP or a domain name served as an effective takedown mechanism. Nowadays, to counter direct communication issues, cybercriminals try to use communication channels that disguise the traffic as benign and cannot be easily blocked, e.g. social networks or use multiple domains to manage infected hosts. In the latter case, the adversary uses a Domain Generation Algorithm (DGA) to periodically generate multiple domains which can be used as rendezvous points to retrieve updates and commands. However, only a few of them are registered. Therefore, the C\&C server can be transferred from one domain to another without losing control of the compromised devices. In addition to the dynamic domain transfers, the lack of proper reporting from domain registrars aggravates the problem, since requests made from law enforcement authorities and security practitioners may receive delayed responses, hindering botnet detection. Therefore, the use of a DGA introduces an asymmetry of cost for the attacker and the defender, as the former has only a minimal cost to register the domains while it is impossible for the latter to block all possible domains.

\subsection{Motivation and Contributions}
\label{sec:motivation}
DGAs come in different flavours depending on how they generate the domain names. The general rule is that they use a pseudo-random generator to create a string that is used as the domain name that the infected hosts would query to reach the C\&C server. To allow the botmaster and bots to generate the same list of domain names, the DGA contains a set of preshared secrets, e.g. the seed. The algorithmically generated domains (AGDs) can be differentiated from benign domains due to a difference in character probability distributions as well as other lexical and statistical features \cite{1,stefanotracking,khaos,Patsakis2019}.

To counter this detection method, some DGAs resort to using random combinations of words which are extracted from predefined dictionaries. Therefore, these DGAs are often referred to as \textit{wordlist-based DGAs}. This way, wordlist-based DGAs bypass many security mechanisms as the generated domains not only have a low entropy, but they also appear to be generated and requested by humans. This is of particular relevance since more than 2/3 of the top domain list contains at least one English word, and around 1/3 are entirely composed by English words \cite{yang2019detecting}. Therefore, distinguishing between benign and malicious word-based AGDs becomes a more challenging task.

In this work, we illustrate that AGDs can be easily identified by the fact that wordlist-based DGAs use a limited dictionary which results in often word repetitions. In this regard, we either exploit the well-known ``\textit{birthday problem}'' and populate custom dictionaries from the Non-existent Internet Domains (NXDomains) that a host queries, or the structure of the domains they produce. Once the queries exceed some quota, we consider that a malware that uses a DGA has infected the host. Despite its simplicity, our method manages to be efficient in terms of both computational effort and detection, allowing it to be easily deployed in existing environments. Notably, the method is more efficient and accurate than the current state of the art, managing to throttle all such DGAs after only a few NXDomain requests without the need for training. Moreover, beyond implementing an additional support layer to detect well-known DGA families, several modules of our method can detect new such DGA families, since they do not depend on training data nor are constrained by domain-specific features. Notably, our methodology is applied to a statistically sound dataset, containing all the known wordlist-based DGA families up to date and 2,189,992 unique domains, which is by far the largest and most complete dataset in the literature of studies for wordlist-based DGAs. This dataset is provided to the research community for reproducing the results and further improvements.

\subsection{Organisation of this work}
The rest of this work is structured as follows. In Section \ref{sec:related}, we present the related work regarding DGAs and detection methods. Then, in Section \ref{sec:methodology}, we discuss the proposed methodology for identifying the operation of a wordlist-based DGA malware, using only network traffic logs. In Section \ref{sec:experiments}, we describe our experimental setup, the datasets we utilise and our results. Afterwards, in Section \ref{sec:discussion}, we discuss the findings of our extensive experiments using our proposed methodology and compare it to the current state of the art. Finally, the article concludes discussing open challenges, future work and summarising our contributions.

\section{Related Work}
\label{sec:related}
Nowadays, malware developers use DGAs, which create a set of AGDs to communicate with C\&C servers, overcoming the drawbacks of static IP addresses \cite{203628,nadji2017still}. In essence, DGAs use a deterministic pseudo-random generator (PRNG) to create a set of domain names \cite{7535098,6175908}. Therefore, the infected devices query a set of domains generated by the DGA till they resolve to a valid IP (i.e. the C\&C server), whose location may also change dynamically. In this regard, blacklisting domains is rendered useless as it implies many practical issues.

According to the literature, there are two main families of DGAs: (i) Random-based DGA methods, which use a PRNG to generate a set of characters to create a domain name, and (ii) Dictionary/Wordlist-based DGA methods, which use a predefined dictionary of existing words to generate such domains and thus, their detection becomes a more challenging task. There also exists a minor subset of DGA families that use valid domains that were previously hacked to hide their C\&C servers (i.e. domain shadowing) \cite{Liu2017} and DGAs that generate domain names that are very similar to existing valid domains or the ones generated by other DGA families \cite{johannesbader} hindering the detection task. Considering the dependency of the pre-shared secret to time, Plohmann et al. \cite{197187} further categorise DGAs to time-independent and deterministic, time-dependent and deterministic, and time-dependent and non-deterministic. Fu et al. \cite{7852496} proposed two DGAs which use hidden Markov models (HMMs) and probabilistic context-free grammars (PCFGs) and tested them on state-of-the-art detection systems. After analysing the outcomes by using metrics such as Kullback-Leibler (KL) distance, Edit distance (ED), and Jaccard index, their results showed that these DGAs hindered the detection rate of such approaches.

In the case of random-based DGA detection, a common practice is to analyse some features of the domain names and their lexical characteristics to determine whether a DGA has generated them \cite{Aviv2011,6151233}. Moreover, auxiliary information such as WHOIS and DNS traffic (e.g. frequent NXDomain responses) is often used to detect abnormal behaviours \cite{Zhou2013DGABasedBD,5762763,1}. Other approaches use machine learning-based techniques and combine the previous information to identify Random-based DGA such as in \cite{5762763,yadav2012,yadavgraph,7163279}.

Nevertheless, many researchers have recently started focusing on the detection of wordlist-based DGAs. In \cite{curtin2018detecting}, authors propose the smashword score, a metric that uses n-gram overlapping combined with information provided from WHOIS lookups to detect AGDs. The WordGraph method \cite{pereira2018dictionary} extracts dictionary information that is embedded in the malware using a graph-based approach, which models repetitions and combinations of domain name strings. In \cite{lison2017automatic}, authors use a machine learning approach based on recurrent neural networks trained using ``familiar'' (i.e. already known) dictionaries to detect wordlist-based AGDs. Similarly, the work presented in \cite{stefanotracking} focuses on AGD classification and characterisation, generating knowledge about the evolving behaviour of botnets. The authors of \cite{Anderson2016} propose a generative adversarial network (GAN), which can learn and bypass classical deep learning detectors. Thereafter, such acquired information is used as feedback to the system to improve the accuracy of the AGD detectors. Neural Networks are also used to classify domain names based on word-level information in \cite{koh2018inline}. More concretely, researchers use ELMo \cite{peters2018deep}, a context-sensitive word embedding, and a classification network that consists of a fully-connected layer with 128 rectified linear units and a logistic regression output layer. In \cite{tong2016method}, the authors propose an improvement of Phoenix botnet detection \cite{stefanotracking} by using a modified Mahalanobis distance metric to perform classification as well as a variant of $k$-means to increase clustering effectiveness. The work described in \cite{18} proposes a short-term memory network (LSTM), which uses raw domain names as features to perform binary classification. Yang et al. proposed a classification based on a set of features such as word correlations, frequency, and part-of-speech tags in \cite{yang2018novel}. Later, they enhanced their detection mechanism by the use of inter-word and inter-domain correlations using semantic analysis \cite{yang2019detecting}. Spooren et al. \cite{Spooren2019} recently showed that their deep learning recurrent neural network is significantly better than classical machine learning approaches. More interestingly though, they showed that one of the dangers of manual feature engineering is that an adversary may adapt her strategy if she knows which features are used in the detection. To this end, they introduce properly crafted DGAs that bypass these classifiers. Berman \cite{berman2019dga} developed a method based on Capsule Networks (CapsNet) to detect AGDs. They compare their method with well-known approaches such as RNNs and CNNs, and the outcomes showed that the accuracy obtained by CapsNet was similar with better performance. Xu et al. \cite{XU201977} proposed the combination of n-gram and a deep CNNs to create an n-gram combined character-based domain classification (n-CBDC) model. Their model runs in an end-to-end way and does not require from domain feature extraction, enhancing its performance. Vinayakumar et al. \cite{Vinayakumar2019} implemented a set of deep learning architectures with Keras Embedding and classical machine learning algorithms to classify DGA families. Their best-reported configuration is obtained when using RNNs with SVM with radial basis function (SVM-RBF). For a detailed overview and classification of methods of how malicious domains can be detected, the interested reader may refer to \cite{Zhauniarovich2018}.

In a recent work \cite{Patsakis2019}, the authors extend the notion of DGAs into a more generic one, namely Resource Identifier Generation Algorithms (RIGA) which allows the use of other protocols beyond DNS. In this regard, authors show how decentralised permanent storage (DPS), although being a useful technology able to enhance a myriad of applications, has some potential drawbacks and exploitable characteristics for armouring a botnet as already exploited in the real world \cite{ipfsstorm}, primarily due to its immutability properties. Therefore, the authors showcase the potential risks and opportunities for malware creators and raise awareness about the symbiotic relationship between DPS and malware campaigns.
Finally, due to recent advances and the widespread use of covert/encrypted communication channels (e.g. DNSCurve, DNS over HTTPS and DNS over TLS) malware creators have an additional layer to hide their communications, rendering traditional DGA detection mechanisms useless. Nevertheless, as shown in \cite{PATSAKIS2020101614}, NXDomain detection can still be performed in such a scenario as well as feature extraction so that DGA families can be further classified with high accuracy.

In addition to the related work analysis, we argue that it is also worth discussing the fact that NXDomain requests may be a result of user typewriting errors. Each human produces different typing patterns depending on the writing surface (e.g. keyboard, smartphone, or larger touch screen surfaces such as tablets) \cite{Findlater2011} and according to its physical and physiological conditions \cite{Andrew88cog,compagno2017don}, which can be used, for instance, to uniquely identify an individual. Nevertheless, typewriting errors are strongly influenced by the language and therefore, exhibit common characteristics regardless of one's typing pattern. The most common typing errors \cite{damerau1964technique,Peterson86let} (more than 80\%) are caused by (i) transposition of two adjacent letters, (ii) adding one extra letter, (iii) one missing letter or (iv) one wrong letter. Therefore, such errors can be corrected backspacing or moving the cursor to the point at which the error occurred and then retyping \cite{Karat99}.

As previously discussed, although typewriting errors may lead to NXDomain with high probability, usually lots of similar domain names to the original are often registered \cite{cartmell2004registering,murphy2003} to avoid homograph attacks \cite{holgers2006cutting}. Moreover, several techniques which aim to overcome homograph attacks can be found in the literature. For example, in \cite{szurdi2014long}, authors designed an accurate typo categorisation framework and found that typosquatting using parked ads and similar monetisation techniques exist for popular domains as well as in the Alexa list. To mitigate this problem, the authors implemented typosquatting blacklists and a browser plugin to prevent mistyping at the user side. In the case of \cite{moore2010measuring}, the authors analyse the main typosquatting issues and monetisation market behind it and recall the effectiveness of several policies and efforts to regulate typosquatting. More recently, novel approaches \cite{226307} developed by the Google Chrome security team, implement suggestions for lookalike URLs, which also prevent typewriting errors. These techniques, added to the fact that most domains are queried after using a search engine, historical data and bookmarks, vastly reduce the number of domains accessed directly through typewriting \cite{BRIN1998107,Aula2005}.

Therefore, we can safely assume that typosquatted domains or homograph attacks represent a marginal percentage or potential danger (compared to DGA queries, which are much more frequent) taking into account the aforementioned prevention and security measures.

\section{Proposed methodology}
\label{sec:methodology}
As already discussed, wordlist-based DGAs have predefined dictionaries that they use to create the possible domains that the malware would try to connect to find the C\&C server. In our methodology, we exploit the fact that this set is often rather constrained, so we expect to have often repetitions of words in the NXDomain requests. The general methodology can be summarised as follows. A monitoring mechanism collects all the NXDomain requests performed by hosts. First, a set of statistical and lexical features of these domains are computed, and next, the domains are split into words, and those words are divided into buckets. The buckets are filled with words either statistically (each word has an individual bucket) or because they fit a specific pattern. If either the features and/or a set of buckets, reach a threshold, an alert is raised.

In what follows, we assume that the monitoring mechanism has a cache that stores the result and would either periodically wipe them after an epoch $T$ or wipe records that are older than $T$. This prevents the mechanism of reporting attacks as a result of, e.g. old typing errors which are stacked over time.

To facilitate the reader, we consider a simple scenario and gradually build on this one to describe our proposal. In our scenario, we have a wordlist-based DGA which selects two words from a dictionary of $n$ words. It adds a separator symbol (e.g. -) between them, and then appends a top-level domain (TLD) from a predefined set. If we assume that $n$ is small, then from the well-known  ``\textit{birthday problem}'' \cite{birthday2} we expect to have a collision, that is a word being repeated, in approximately $\sqrt{n}$ domain name generations. More precisely, the latter is expected to happen with 50\% probability.

Setting the threshold of repetitions too low, e.g. 2, it is evident that may lead to many false positives as a user may have mistyped a domain name. This small amount of false positives is an acceptable trade-off in the event of human errors; as described in Section \ref{sec:related}. Although typewriting attacks are much less frequent than DGA-based malware, as discussed in Section \ref{sec:related}, one may set the threshold for repetitions of words in domain names higher to allow for some grace for typos. In what follows, we denote this threshold as $t$.

Generalising the above, one DGA may have $k$ dictionaries and generate each fragment of the domain name by selecting a word from each dictionary. Therefore, we may formalise our problem as follows:

\noindent\textbf{Problem setting:}
\textit{Let us assume that a DGA has $k$ dictionaries $d_i, \: i \in \{1,...,k\}$. The DGA uses words (denoted as $w$) to create domain names of the form $dom$ such that:}

\[
    dom=w_1||w_2||...||w_k, \: w_i\in d_i, \: \forall i\in\{1,2,3,...,k\}.
\]

\textit{That is, $dom$ is the ordered concatenation ($||$) of $k$ words by randomly selecting one word from each dictionary and putting them in the same order as their dictionaries. Find the probability $p$ of having at least one word from any of the dictionaries being selected at least $t$ times, for $t$ constant.}

It is clear that the case of having one dictionary ($k=1$) and requesting one collision ($t=2$) is the well-known birthday problem. Levin \cite{levin1981representation} and Diaconis and Mosteller \cite{diaconis1989methods} have thoroughly studied the birthday problem and its extensions. Based on their proofs, we have that the probability $p$ of having $t$ collisions on a dictionary of $L$ words in $n$ trials is subject to the following approximation:
\[
\frac{ne^{-n/Lt}}{\left(1-\frac{N}{L(t+1)}\right)^{1/t}}\approx \left(L^{t-1}t!\log\left(\frac{1}{1-p}\right)\right)^{1/t}.
\]

Therefore, solving for p we have that:
\[
p\approx
1-\exp \left(-\frac{L^{1-t} \left(n e^{-\frac{n}{Lt}}
\left(1-\frac{n}{L(t+1)}\right)^{-1/t}\right)^t}{t!}\right).
\]

From the latter approximation, and the fact that in the generic wordlist-based DGAs, discussed above, the collisions may affect different dictionaries are independent; one can compute the probability of a $t$-collision as follows:
\[
P(t-collision)=\sum_{i=1}^kp_i,
\]
where:
\[
p_i=1-\exp \left(-\frac{L_i^{1-t} \left(n e^{-\frac{n}{L_it}}
\left(1-\frac{n}{L_i(t+1)}\right)^{-1/t}\right)^t}{t!}\right).
\]
and $L_i$ denotes the length of dictionary $d_i$.

Going a step further, let us assume that we have captured an NXDomain name request from a host. We may split the name to see the pattern of the domain name in terms of structure. For instance, the domain name consists of 2 or 3 words with a total length above $M$ characters. Note that DGAs tend to have rather long domain names to, e.g. maximise the chances of purchasing the domains they want and avoid compromised machines contacting existing domains. Therefore, due to the way wordlist-based DGAs work, requests to NXDomains which have long names or with a specific amount of concatenated words may imply the existence of an infected machine. More precisely, wordlist-based DGAs often have a static template of the form, e.g. \texttt{w$_1$||w$_2$.TLD} (see DGAs like \texttt{nymaim}, \texttt{pizd}, and \texttt{suppobox}) or \texttt{w$_1$||w$_2$||...||w$_m$.TLD} so that:
\[
|w_1|+|w_2|+...+|w_m|>M,
\]
which contents would change according to each DGA. As it is shown in Section \ref{sec:experiments}, typical user DNS queries do not follow this template.

\begin{figure*}
    \centering
    \includegraphics[width=0.8\textwidth]{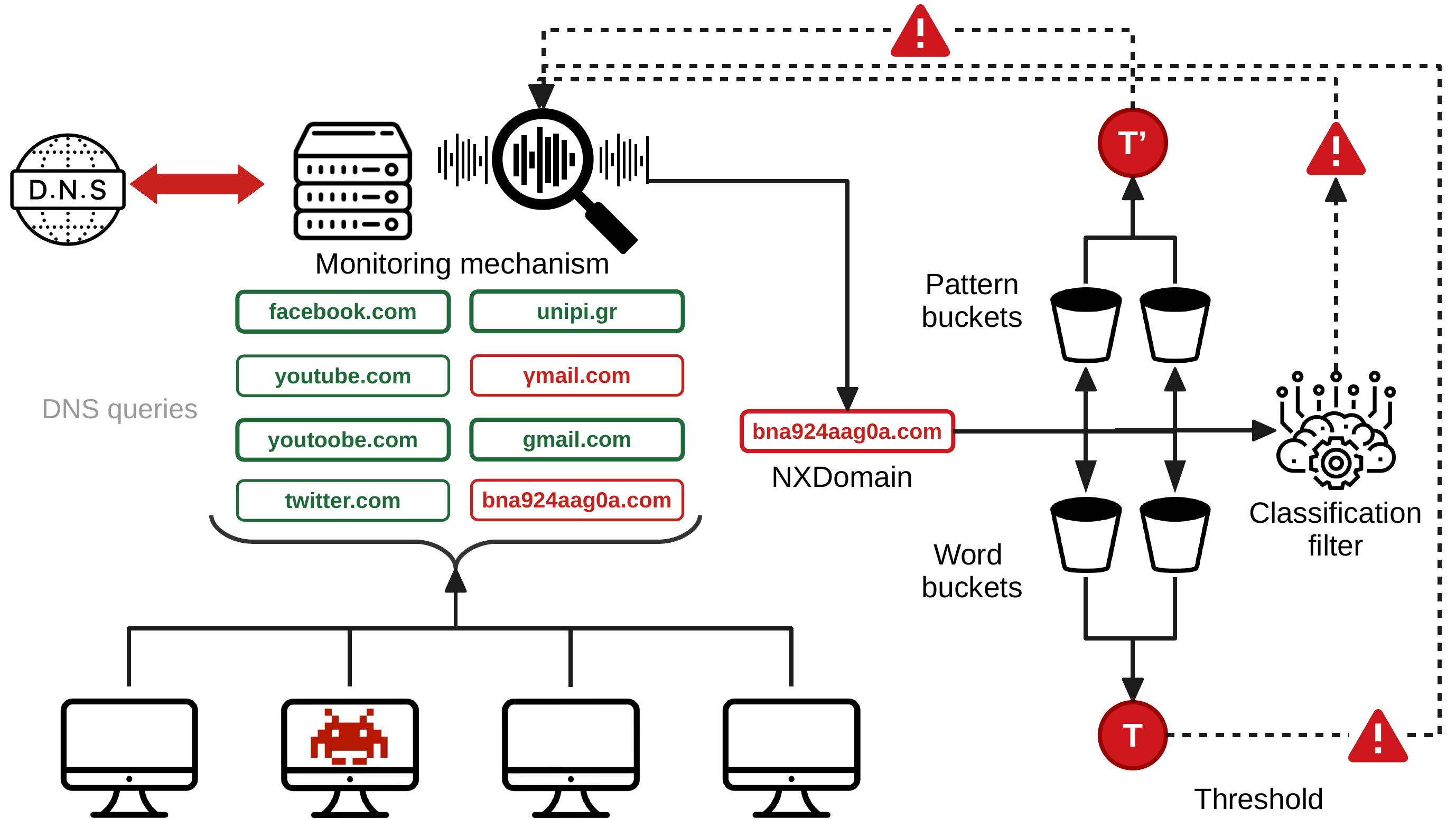}
    \caption{An illustration of the proposed methodology. Network devices (see below) perform DNS queries which are captured by the monitoring mechanism. Our filters then process all NXDomain requests, and if they exceed specific thresholds, the corresponding alerts are raised. }
    \label{fig:proposed_methodology}
\end{figure*}

In addition to the word and pattern filters, we apply another filter, namely a classification filter, which works in real-time and analyses the statistical and lexical features of an SLD to determine whether it has been created by a DGA or not. Therefore, the proposed methodology is a three-layer filter for the collected NXDomains. By default, DNS queries are performed in cleartext format and can be captured by any network device that is hosted in the same network. Therefore, a network monitoring device can passively collect all DNS queries performed in the network and submit the NXDomains for processing in our three-filter mechanism.
The proposed methodology is illustrated in Figure \ref{fig:proposed_methodology}.

First, the network monitoring mechanism collects all the DNS queries and logs all NXDomain requests. Each such request is analysed by the three filters to check whether a threshold is exceeded. To do so, we first remove the TLD of the domain request, and we keep only the SLD which is analysed. The bulk of the literature studies SLDs only, as TLDs offer little information and they are well-known and fixed. Third level domains, since any anomaly in them, can be easily detected and handled controlled by the owner of the SLD are not considered in the related work. Next, the classification filter receives as an input the SLD and performs a classification based on its features, which outcome is either such SLD has been created by a DGA or not. The word filter receives as an input the SLD, splits it into words and sets a counter for each word that appears in such analysis. Once the counter of a word exceeds a given threshold $T$, a warning is issued. In parallel, the SLD is analysed for structural patterns, e.g. number of words, total length etc. as later discussed in Section \ref{sec:experiments}. Again, if the counter that keeps track of repetitions of these patterns exceeds a given threshold $T'$, a warning is raised.

As it will be discussed in Section \ref{sec:discussion}, the three-layer approach, which analyses the domain names in parallel, manages to provide high reliability, accuracy, and robustness. Moreover, due to the template that such AGDs are expected to have, our methodology is generic enough to counter other and non-reported wordlist-based DGAs.

\section{Experiments}
\label{sec:experiments}

In this section, first, we describe the setup and methodology of our experiments. Next, we use the implementation of several DGAs, as provided by researchers\footnote{\url{https://github.com/baderj/domain_generation_algorithms},
\url{https://github.com/andrewaeva/DGA}}\textsuperscript{,}\footnote{\url{https://github.com/ynvb/ExplosiveScripts}}  who have reversed engineered the corresponding malware that embeds them, the real-world captures of the DGArchive \cite{197187}, and two DGAs that were designed to bypass machine learning algorithms \cite{Spooren2019}. More concretely, we study the following DGA families: \texttt{beebone}, \texttt{banjori}, \texttt{gozi}, \texttt{matsnu}, \texttt{nymaim2}, \texttt{pizd}, \texttt{rovnix}, \texttt{suppobox} and \texttt{volatilecedar}. Apart from these families, we include in our dataset two arithmetic-based DGAs  \texttt{beebone} and  \texttt{banjori} and one permutation-based (\texttt{volatilecedar}) to see the effectiveness of our method in other families of DGAs. In addition, we included several adversarial DGA families. The first two are \texttt{khaos} \cite{khaos} and \texttt{charbot} \cite{8756038}. The other two families have been retrieved from \cite{Spooren2019}, namely \texttt{decept} and \texttt{decept2}, all of which were crafted to bypass machine learning-based mechanisms. Table \ref{tbl:sample} provides an overview of our dataset.
To this end, the table presents how many samples each DGA has and how many seeds. Moreover, we provide ten AGDs from each DGA in our dataset to facilitate the reader in understanding the AGDs of each DGA produces.

The underlying dictionaries vary in length, origin, and amount for each DGA. More precisely, \texttt{matsnu} contains two dictionaries, one for verbs (878) and one for nouns (1008). Depending on the seed its AGDs will either always start from a verb or a noun. Then, using the seed for its PRNG, \texttt{matsnu} selects one word from each dictionary iteratively from each dictionary, until the length of the domain exceeds 24 characters.
\texttt{nymaim2} uses two dictionaries, one for the first word (2450 words) and one for the second one (4387 words) which are concatenated either directly without any separator or with the ``-'' character. \texttt{pizd} uses by default a wordlist containing 384 words and uses a PRNG to select two words and concatenate them to generate an AGD. \texttt{rovnix} uses as its source the US Declaration of Independence. The dictionary contains all the alphanumeric words of the document. To construct an AGD, \texttt{rovnix} selects words according to a PRNG until the selected words exceed the 20 characters. \texttt{gozi} is a variant of \texttt{rovnix}. Its various dictionaries originate from various public domain documents that are unlikely to be moved to another location, e.g. Request for Comments pages and the GNU Lesser General Public License. From these documents, \texttt{gozi} splits the documents according to stop characters (spaces, commas, etc.) and selects the words with at least three characters that contain only letters. From this wordlist, the PRNG selects random words that are concatenated so that the resulting AGD contains between 12 and 23 characters. \texttt{suppobox} in the three seeds that have been identified so far has a dictionary of 384 words from which it selects two random words and concatenates them.

In our dataset, we have also added the top 1 million websites from Alexa. The reason for including this is to have some ground truth of benign traffic and illustrate the domains that an actual user would use is significantly different from the ones that would be derived from a DGA, even if they are made to do so. It should be noted here that the Alexa top 1 million dataset contains web pages and not domains. Therefore, there are repetitions of domains, e.g. blogspot. Moreover, there are several Internationalized domain names (IDN)\footnote{https://www.icann.org/resources/pages/idn-2012-02-25-en} (i.e. domains that start with the characters \texttt{xn--}) therefore, the domain name cannot contain any word. We opted to remove the latter domains and to keep each domain once. Note that up to now, there is no DGAs using IDN domain names. Therefore, the Alexa dataset consists of 915,994 unique domains.

Finally, we have an additional real-world dataset consisting of NXDomain requests as collected from our institution. The dataset was provided by our network department, containing only the NXDomain requests that were performed; therefore, the data were completely anonymised. The data collection period was for a week, and it involved all the NXDomain requests that were performed in the whole network of the University of Piraeus. From these domains, we kept each domain once.
The final dataset that from now on, we will refer to as \texttt{unipi} consists of 3547 domains. The use of this dataset in our experiments enables us to evaluate our method against real-world NXDomain requests. It should be noted that some of these requests are malicious, however, this will be discussed in detail in Section \ref{sec:discussion}.

\begin{sidewaystable}[h]
\centering
\scriptsize
\rowcolors{2}{gray!25}{white}
\begin{tabular}{lp{1in}rcp{3.5in}}
\toprule
\textbf{DGA} & \textbf{Type} & \textbf{\# of samples} & \textbf{Seeds} &\textbf{Sample domains} \\
\midrule
alexa & Benign & 915,993 &-  & google.com, youtube.com, tmall.com, baidu.com, qq.com \\
unipi & Benign \& Malicious & 3,574  &-&  openstrreetmaps.gr, wheagoogle.com, naftempoirik.gr, j4ng4uhpx51v.com, 3837avw-2iay7bstddjg0b.com\\
beebone & Arithmetic & 42& 2  &backdates0.com, backdates0.org, backdates0.net, backdates0.biz, backdates0.info\\
banjori & Arithmetic & 439,068 & 35  & andersensinaix.com, xjsrrsensinaix.com, hlrfrsensinaix.com, fnosrsensinaix.com, qcwcrsensinaix.com\\
charbot & ML & 99,785 & 1 & profumrdalzorno, ppgvoic7dfcolor, bkeasnnexum, lutt6r \\
decept & ML & 149,288 & 1 & ytrbegofitr8b, rithundiat, tarenoth200qyumurop, hatoranfa, xunrvrstbe\\
decept2 & ML & 149,638 & 1  &tafersickir, pblogalmarportran-f, martapord, joedavingsbiosk, prialtions\\
khaos & ML & 9,772 &1& ycoentokai hsvo5yoanl, proytni, cessit-tfabilition, oail,
\\
gozi & Wordlist & 136,306 & 104  &penarumsalpaesthodie.com, quodquibusfulminatcur.com, defunctorumnullamrelaxat.com, veniarumcuramhabet.com, nisisacerdotipapefalse.com\\
matsnu & Wordlist & 31,096& 3 & starsendbottomhabitshake.com, causeirongroundnettellstart.com, cultureexploredogdistrict.com, sizeprogrambillsaypointpot.com, tourmentionboneconcertadmire.com\\
nymaim2 & Wordlist & 44,339 & 1 &squirting-eight.net, unitsedgar.net, sexuality-giant.net, utilitiespour.ki, vermontfeatures.ad\\
pizd & Wordlist & 15,286 & 1  &aboveshare.net, alreadyshare.net, actionprobable.net, althoughprobable.net, actionseveral.net\\
rovnix & Wordlist & 99,747 & 1 &   theirtheandaloneinto.com, thathistoryformertrial.com, tothelayingthatarefor.com, definebritainhasforhe.com, tosecureonweestablishment.com\\
suppobox & Wordlist & 95,560 &3  &possibleshake.net, mountainshare.net, possibleshare.net, perhapsnearly.net, windownearly.net\\
volatilecedar & Permutation & 498& 2 &  deotntexplorer.info, doetntexplorer.info, dotentexplorer.info, dotnetexplorer.info, dotnteexplorer.info
\\
\textbf{Total} & & \textbf{2,189,992}& \\\bottomrule
\end{tabular}
\caption{Overview of our dataset.}
\label{tbl:sample}
\end{sidewaystable}

Using real data from the DGArchive \cite{dietrich2011botnets} and the reversed code, we collect the set of the domain names they create. The total AGDs are 1,270,425 and adding the benign domains from Alexa and unipi; we end up with a dataset of 2,189,992 unique domains. The dataset and additional information can be found in \cite{constantinos_patsakis_2020_4010620}.
Each domain in the dataset is processed to extract a set of features and the words that were concatenated to generate them as well. The latter is achieved with the use of Wordninja\footnote{https://github.com/keredson/wordninja}, a natural language processing (NLP) method that probabilistically splits concatenated words based on English Wikipedia uni-gram frequencies.

In all filters, we assume that the network monitoring device intercepts all the NXDomain requests from each device and analyses them for each one individually. For the sake of clarity, we assume that the network monitoring device gets as input a stream of NXDomain requests from only one device. Moreover, we assume that the device is infected by only one malware with DGA capabilities, and therefore the malware uses one seed. In our experiments, this practically means that each one of them has to be executed per DGA family and using one seed at a time. We argue that this is strategy is correct as the malware will not manage to connect directly to the C\&C server and several connections would be attempted before the malware manages to connect and receives the command to, e.g. use a different dictionary/seed. Note that in our DGA dataset, the different seed in several cases translates to the use of another dictionary. Nevertheless, our classification filter is able to manage any configuration regardless of the seed and the amount of DGAs infecting a machine, as described below.

We created a classification filter that will analyse all the domain names queried as follows. First, we cache the domain name for analysis, and we compute a set of statistical, lexical (including a 2-Chain Markov model for English grams\footnote{https://github.com/rrenaud/Gibberish-Detector}) and entropy features. The set of features computed for each domain name is depicted in Table \ref{tab:features}. After computing such features and using a classification scheme, we are able to determine, with high accuracy, whether the domain name is benign or not without the need for external information or waiting for the domain name resolution response.


\begin{table}[!ht]
\centering
\scriptsize
\caption{Features used in our approach and the corresponding description.}
\label{tab:features}
\begin{tabular}{>{\bfseries}cp{1in}p{2.8in}}
\toprule
\textbf{Feature Set} & \textbf{Notation} & \textbf{Description}  \\
\midrule
\multirow{7}{*}{Alphanumeric Sequences} & $Dom$ & Domain without TLD   \\   
&$Dom-D$  & $Dom$ without the digits  \\   
&$Dom-3G$  & Set of 3-grams of $Dom$  \\
&$Dom-4G$  & Set of 4-grams of $Dom$     \\
&$Dom-5G$  & Set of 5-grams of $Dom$   \\   %
&$Dom-WS$ & Domain concatenated words with spaces  \\  
&$Dom-WDS$ & $Dom-D$ concatenated words with spaces  \\
&$Dom-W2$ & Domain concatenated words of length $>$ 2  \\  
&$Dom-W3$ & Domain concatenated words of length $>$ 3  \\   
\midrule
\multirow{5}{*}{Statistical Attributes}& $L-HEX$ & The domain name is represented with hexadecimal characters \\
& $L-LEN$ & The length of $Dom$  \\
& $L-DIG$ & The number of digits in $Dom$  \\
& $L-CON-MAX$ & The maximum number of consecutive consonants $Dom$\\
& $L-W2$ & Number of words with more than 2 characters in $Dom$\\
& $L-W3$ & Number of words with more than 3 characters in $Dom$ \\ 
 \midrule
 \multirow{10}{*}{Ratios and Lexical Attributes} & $R-CON-VOW$ & Ratio of consonants and vowels of  $Dom$ \\
&$R-Dom-3G$  & Ratio of benign grams in $Dom-3G$  \\   %
&$R-Dom-4G$  & Ratio of benign grams in $Dom-4G$    \\  %
&$R-Dom-5G$  & Ratio of benign grams in $Dom-5G$  \\   %
&$M2-Dom-WS$  & 2-Chain Markov English grams applied to $Dom-WS$ \\
&$M2-Dom-WDS$  & 2-Chain Markov English grams applied $Dom-WDS$ \\
&$R-WS-LEN$ & $Dom-WS$ divided by $L-LEN$ \\
&$R-WDS-LEN$ & $Dom-WDS$ divided by $L-LEN$  \\
&$R-W2-LEN$ & $Dom-W2$ divided by $L-LEN$ \\
&$R-W3-LEN$ & $Dom-W3$ divided by $L-LEN$  \\
\midrule
\multirow{4}{*}{Entropy}&$E-Dom$  & Entropy of  $Dom$ \\
&$E-Dom-WS$  & Entropy of $Dom-WS$ \\
&$E-Dom-WDS$  & Entropy of $Dom-WDS$ \\
&$E-Dom-W2$  &  Entropy of $Dom-W2$ \\
&$E-Dom-W3$  &  Entropy of $Dom-W3$  \\
\bottomrule
\end{tabular}
\end{table}

We employ a Random Forest (RF) model, which is a non-parametric ensemble classifier. RF is widely used in state of the art, achieving outstanding performance results on DGA classification tasks~\cite{Anderson2016,Spooren2019}. The hyperparameters of the RF algorithm were tuned with grid search, to maximise classification performance in the task of distinguishing between benign and malicious domains in a subset of our dataset. We found that best performance is achieved using an ensemble of 100 decision trees with unlimited depth and bootstrap aggregation (bagging), where each new tree is fitted from a bootstrap sample of the training data~\cite{breiman1996bagging}. In all experiments, we employed 10-fold cross-validation to get an unbiased estimate of the classification accuracy. Moreover, we employed random sampling without replacement for AGDs and repeated the experiments 100 times with the previous setup, to guarantee statistically sound outcomes. Table \ref{tab:binary_stats} shows the outcomes of our classification. All our experiments were performed on a system equipped with an NVIDIA TITAN Xp PG611-c00 to boost the performance, while we utilised the implementations of the \texttt{scikit-learn}\footnote{\url{https://scikit-learn.org}} library. We evaluate the performance of the trained classifiers using the standard classifications metrics of Precision, Recall, and $F_1$-score.

\begin{table}[h]
\setlength{\tabcolsep}{3pt}
\def\arraystretch{1.1}
\centering
\small
\caption{Performance measures for the binary classification (in percentage) averaged over 100 experiments.}
\label{tab:binary_stats}
\begin{tabular}{>{\bfseries}lcccc}
\hline
\textbf{Class}       & \textbf{Prec.}  & \textbf{Recall} & \textbf{$F_1$}   &\textbf{$\sigma_{F_1}$} \\
\midrule
khaos         & 89.76 & 83.89 & 86.73 & 0.11 \\
banjori       & 99.94 & 99.54 & 99.74 & 0.03 \\
beebone       & 99.30 & 100   & 99.65 & 0.57 \\
charbot       & 99.65 & 98.22 & 98.93 & 0.09 \\
decept     & 90.68 & 87.97 & 89.31 & 0.15 \\
decept2    & 86.89 & 84.55 & 85.70 & 0.32 \\
gozi          & 92.81 & 92.93 & 92.87 & 0.20 \\
matsnu        & 87.30 & 92.75 & 89.94 & 0.15 \\
nymaim2       & 88.14 & 91.30 & 89.69 & 0.34 \\
pizd          & 91.29 & 96.05 & 93.61 & 0.44 \\
rovnix        & 97.76 & 99.62 & 98.68 & 0.08 \\
suppobox      & 85.68 & 91.48 & 88.49 & 0.25 \\
volatilecedar & 99.47 & 97.59 & 98.52 & 0.18\\
\bottomrule
\end{tabular}
\end{table}

As it can be observed, we can distinguish, with detection rates ranging from 85.70\% to 99.74\%, between malicious and benign domains for the tested families. The most difficult families to distinguish were \texttt{decept2} and \texttt{khaos} (i.e. according to the $F_1$ metric). Nevertheless, as we will discuss in Section 5, the outcomes obtained for adversarial families outperform the current state of the art. Moreover, the $\sigma_{F_1}$ values remain low, showcasing the statistical relevance of the outcomes. Note that capturing with high accuracy all the word-based DGAs with the same feature set evinces the efficacy of our approach. With such a classification filter, our approach is able to discern between malicious and benign domains at each iteration. Moreover, when the corresponding thresholds of either the word or the pattern buckets are reached, the system will trigger an alert, ensuring AGD detection. We measured several statistics during our classification pattern experiments, such as the time required to compute the features and the prediction time in the binary setting with our dataset. The average time required to compute all the features for an SLD is 0.54 ms, while the prediction time is 1.05 ms on average. Therefore, this filter is suitable for real-time AGD detection, even in environments with a high volume of traffic.

\begin{figure*}[!ht]
    \centering
    \begin{subfigure}{0.49\textwidth}
        \centering
        \includegraphics[width=.9\columnwidth]{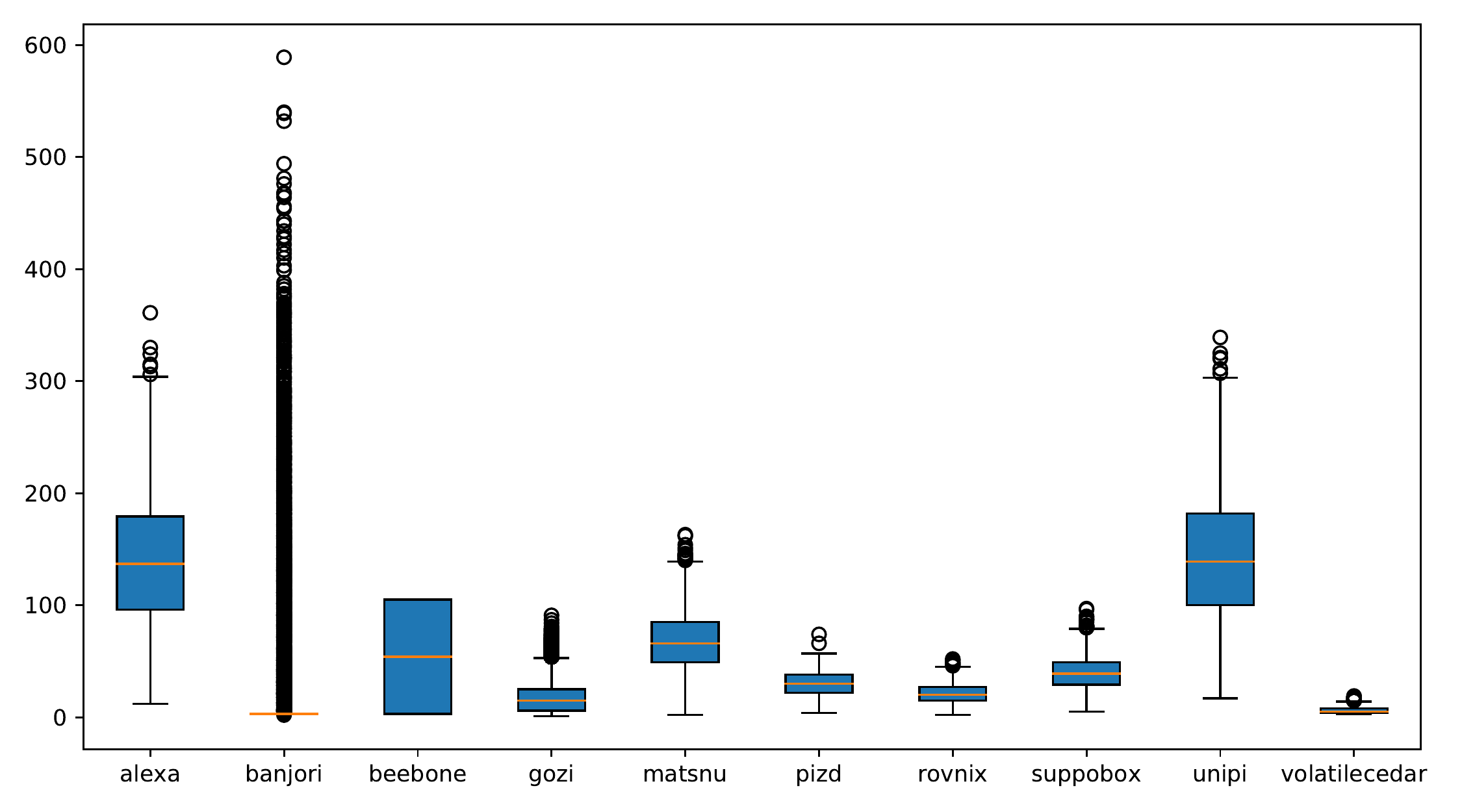}
        \caption{3 strikes.}
    \end{subfigure}~
        \begin{subfigure}{0.49\textwidth}
        \centering
        \includegraphics[width=.9\columnwidth]{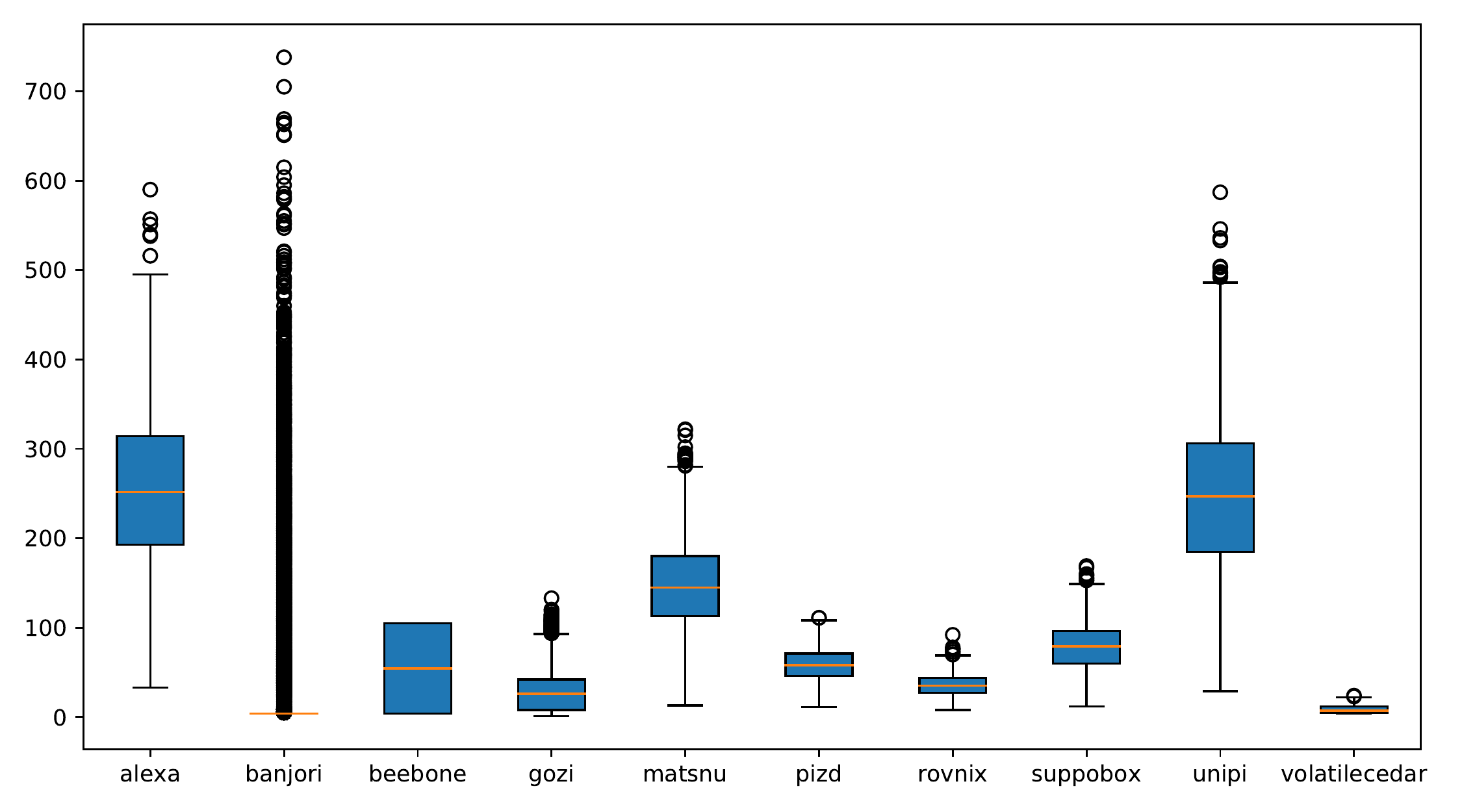}
        \caption{4 strikes.}
    \end{subfigure}

    \begin{subfigure}{0.49\textwidth}
        \centering
        \includegraphics[width=.9\columnwidth]{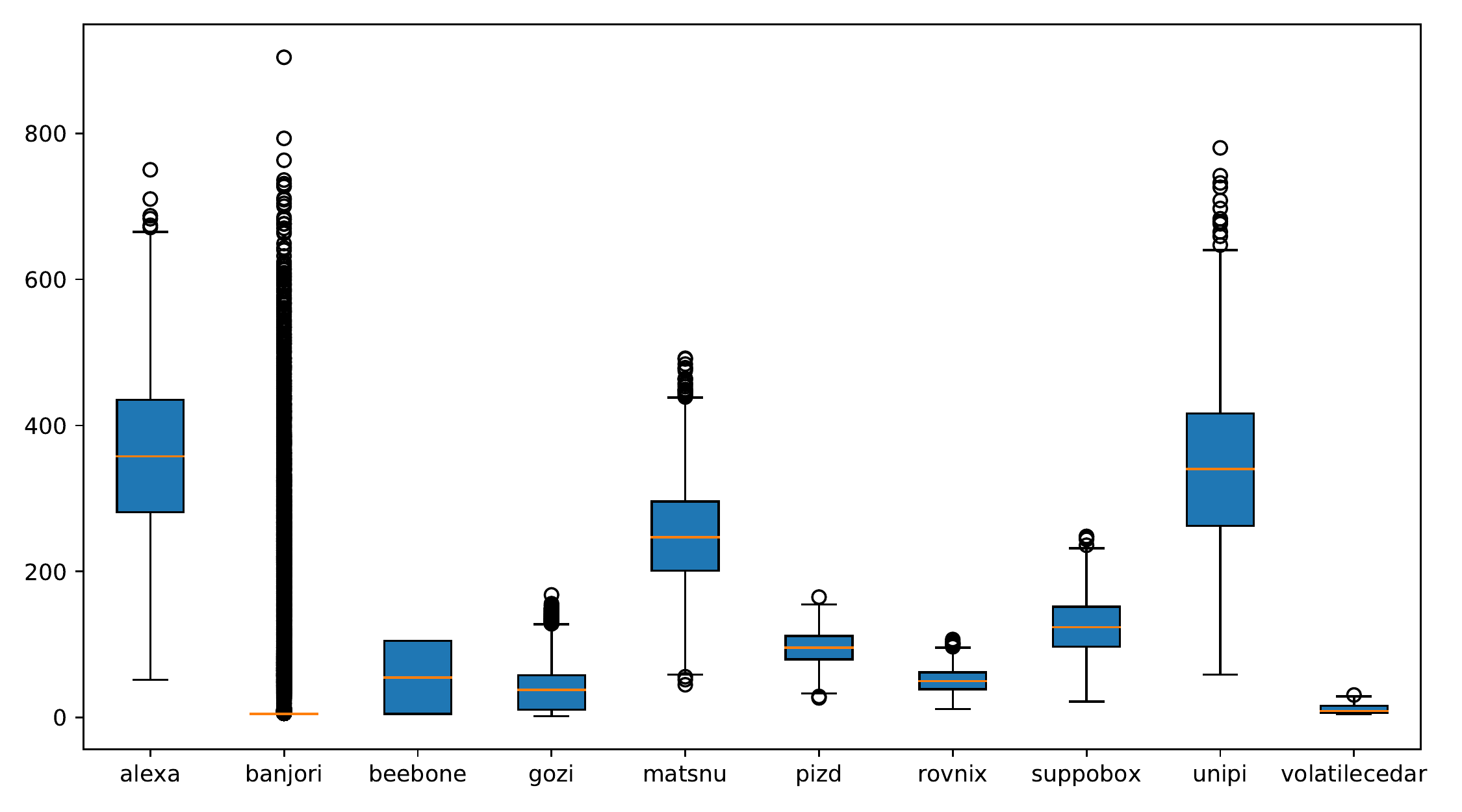}
        \caption{5 strikes.}
    \end{subfigure}~
    \begin{subfigure}{0.49\textwidth}
        \centering
        \includegraphics[width=.9\columnwidth]{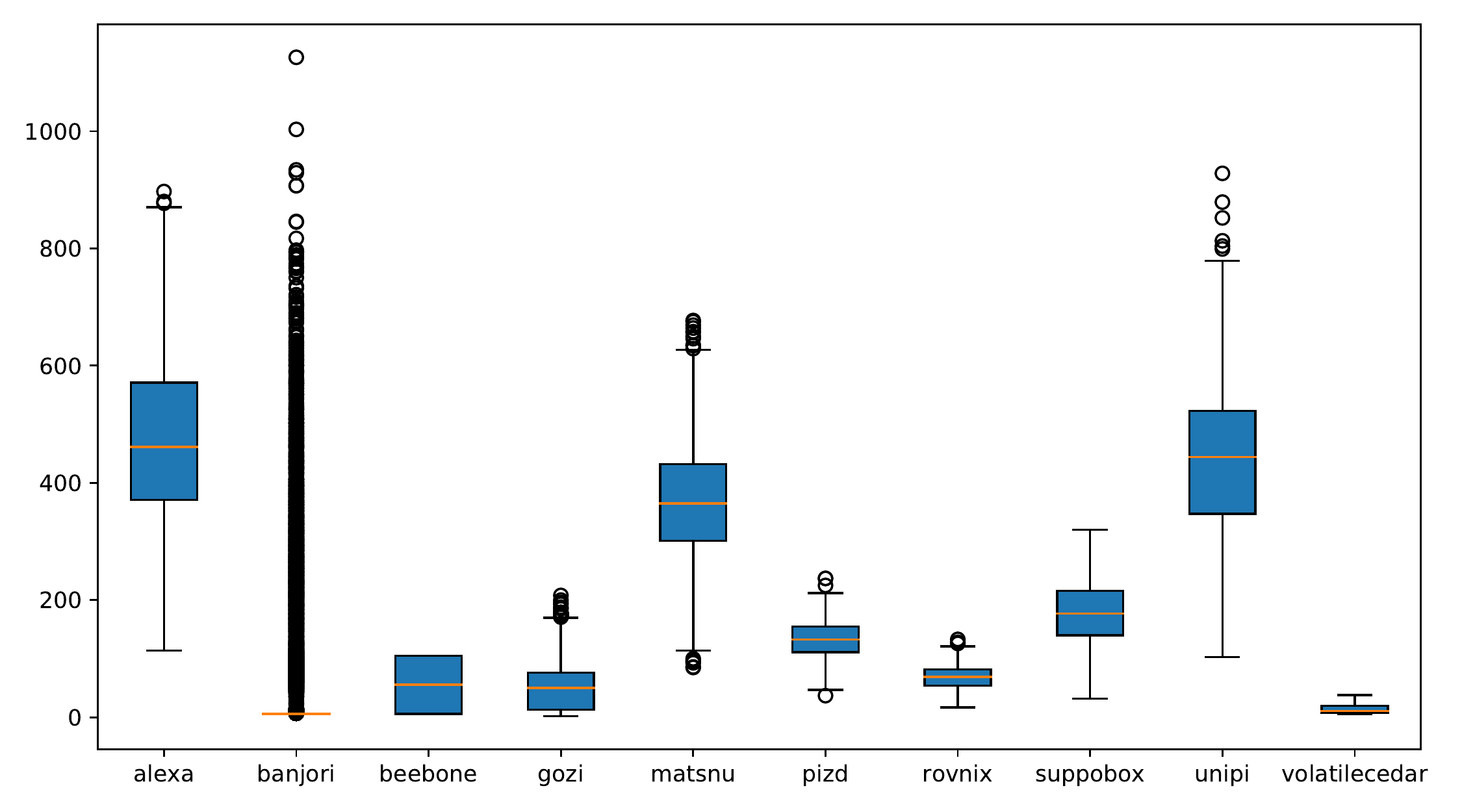}
        \caption{6 strikes.}
    \end{subfigure}

    \begin{subfigure}{0.49\textwidth}
        \centering
        \includegraphics[width=.9\columnwidth]{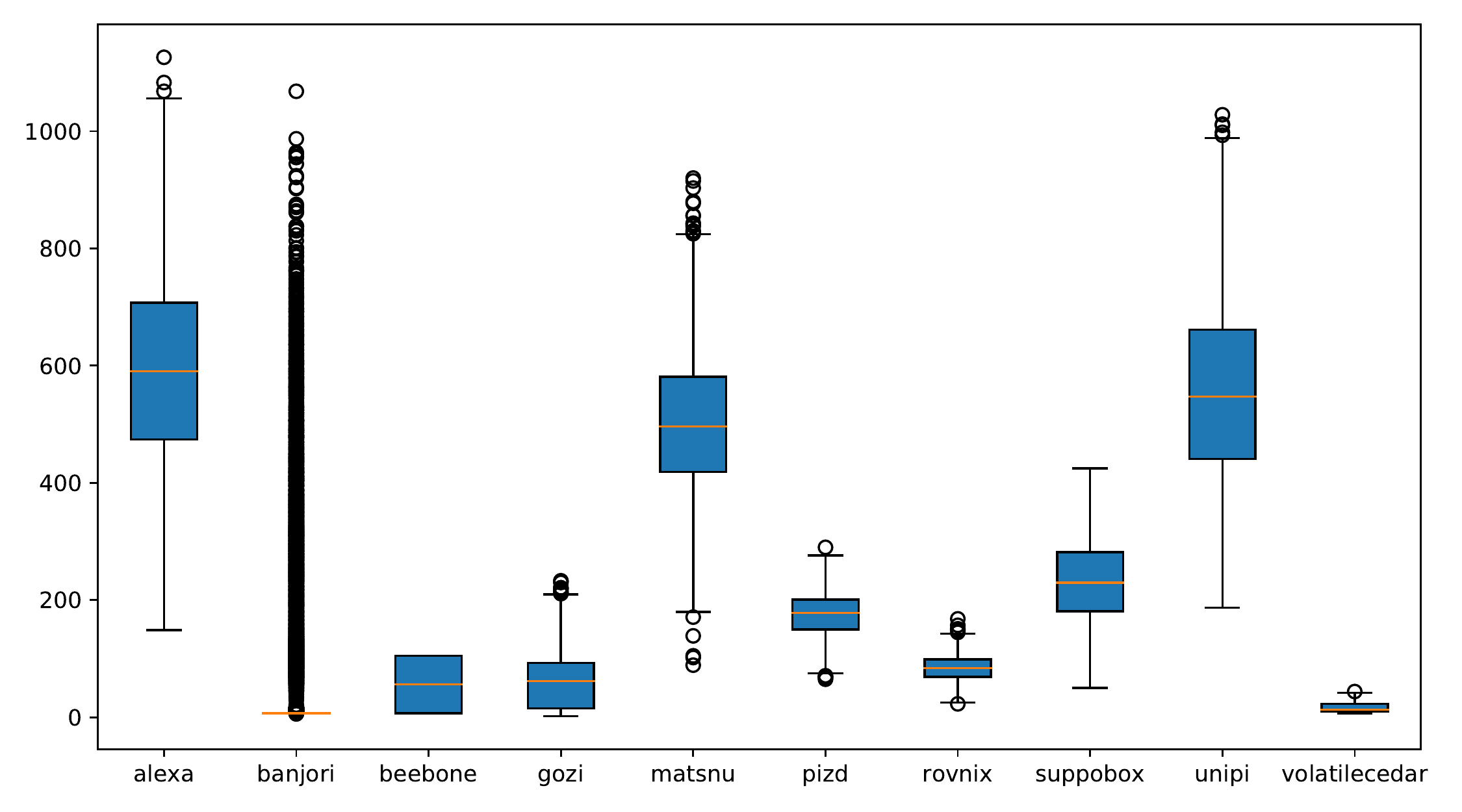}
        \caption{7 strikes.}
    \end{subfigure}~
    \begin{subfigure}{0.49\textwidth}
        \centering
        \includegraphics[width=.9\columnwidth]{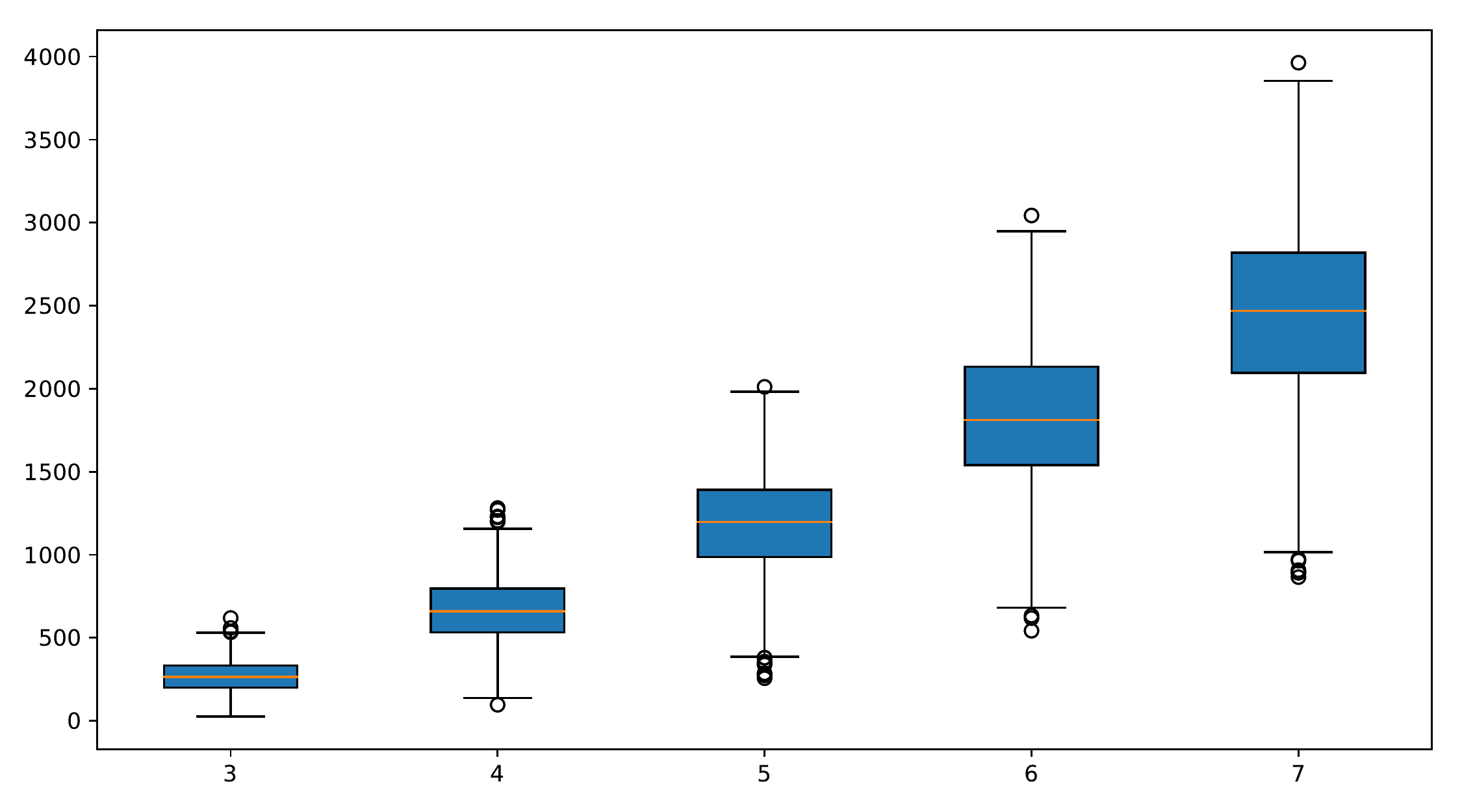}
        \caption{The statistics of \texttt{nymaim2} for different stikes.}
        \label{fig:nymaim2detail}
    \end{subfigure}

\caption{Number of domains, generated by each DGA family, that are needed to reach the strike threshold.}
\label{fig:strikes}
\end{figure*}

In the next set of experiments, we test the frequency of word collisions on specific thresholds. More precisely, we split each domain name in words and record their occurrences if their length is more than three letters to avoid stop words, articles, pronouns etc. Based on a threshold of how many occurrences we expect from an AGD during an epoch, we monitor all NXDomain requests and raise an alert when the threshold is reached. To provide better insight on these results, more to than simply reporting the first time that an NXDomain query is performed for each AGD and seed in our dataset, we shuffled them and made the same measurement 1,000 times. In Figure \ref{fig:strikes}, we illustrate the results for different threshold levels, which range from 3 to 7. By shuffling the domains that a DGA generates, we test the DGA with different possible seeds, far more than the ones we originally had. Since the use of different seed is a common practice in malware, we may study how our methodology performs in different settings, and show that it is generic enough to be used in various configurations. Although this threshold can be modified, we claim that three unrelated NXDomain queries that contain the same word are likely not to be generated by a human, according to Section \ref{sec:related} as well as discussed in our problem setting. This hypothesis is confirmed from our first experiment, which shows that our word filter is able to detect the domains generated by each family of DGA accurately. It should be noted that the real-world \texttt{unipi} dataset exhibits very similar characteristics to the Alexa dataset. Despite the existence of DGA domains, most of the domains are typed, or more precisely, are mistyped by humans, so they have similar statistical characteristics with the Alexa dataset.

\begin{figure*}[!ht]
    \centering
    \begin{subfigure}{0.49\textwidth}
        \centering
        \includegraphics[width=.9\columnwidth]{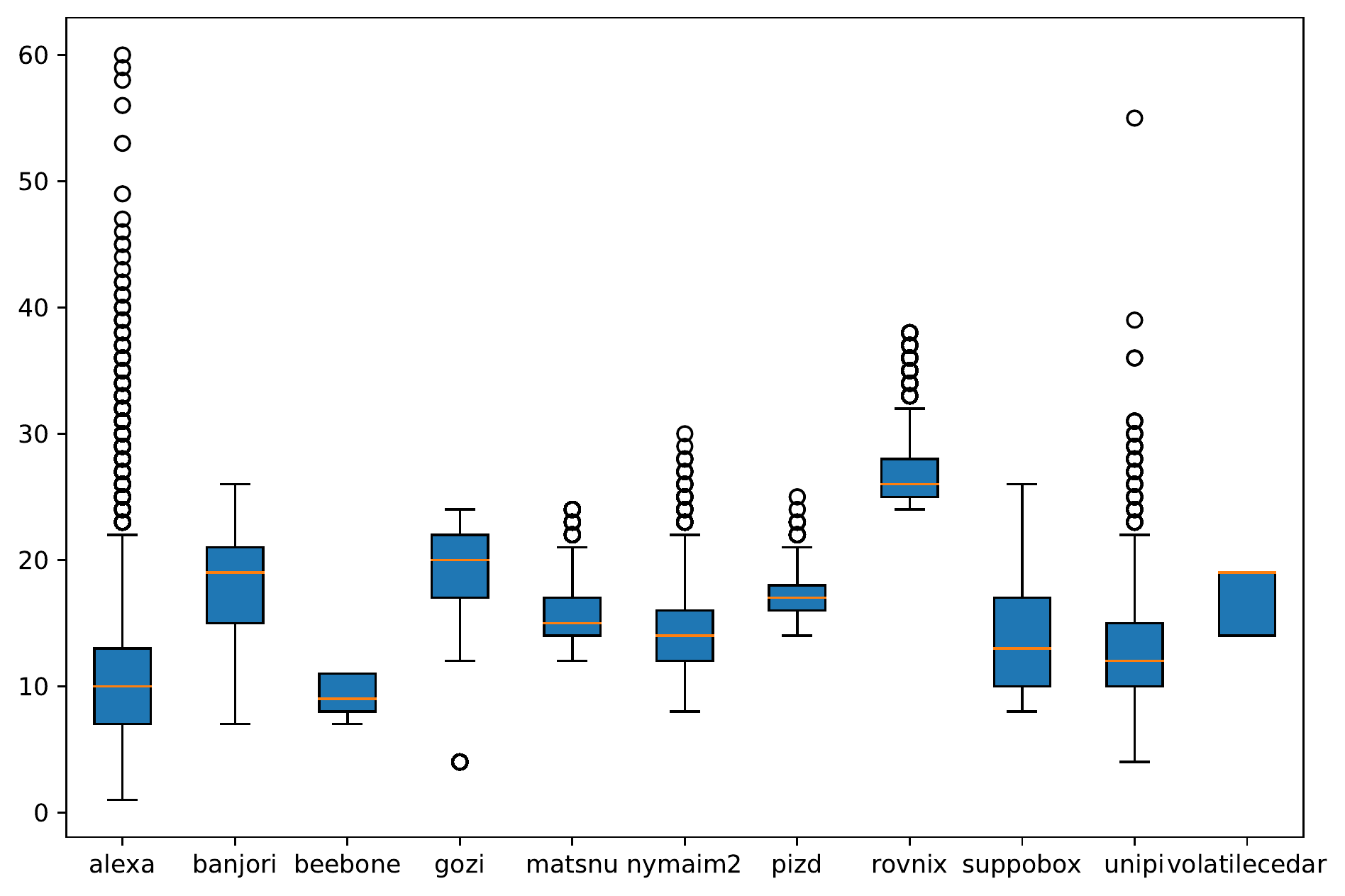}
        \caption{Average letters per domain.}
    \end{subfigure}~
        \begin{subfigure}{0.49\textwidth}
        \centering
        \includegraphics[width=.9\columnwidth]{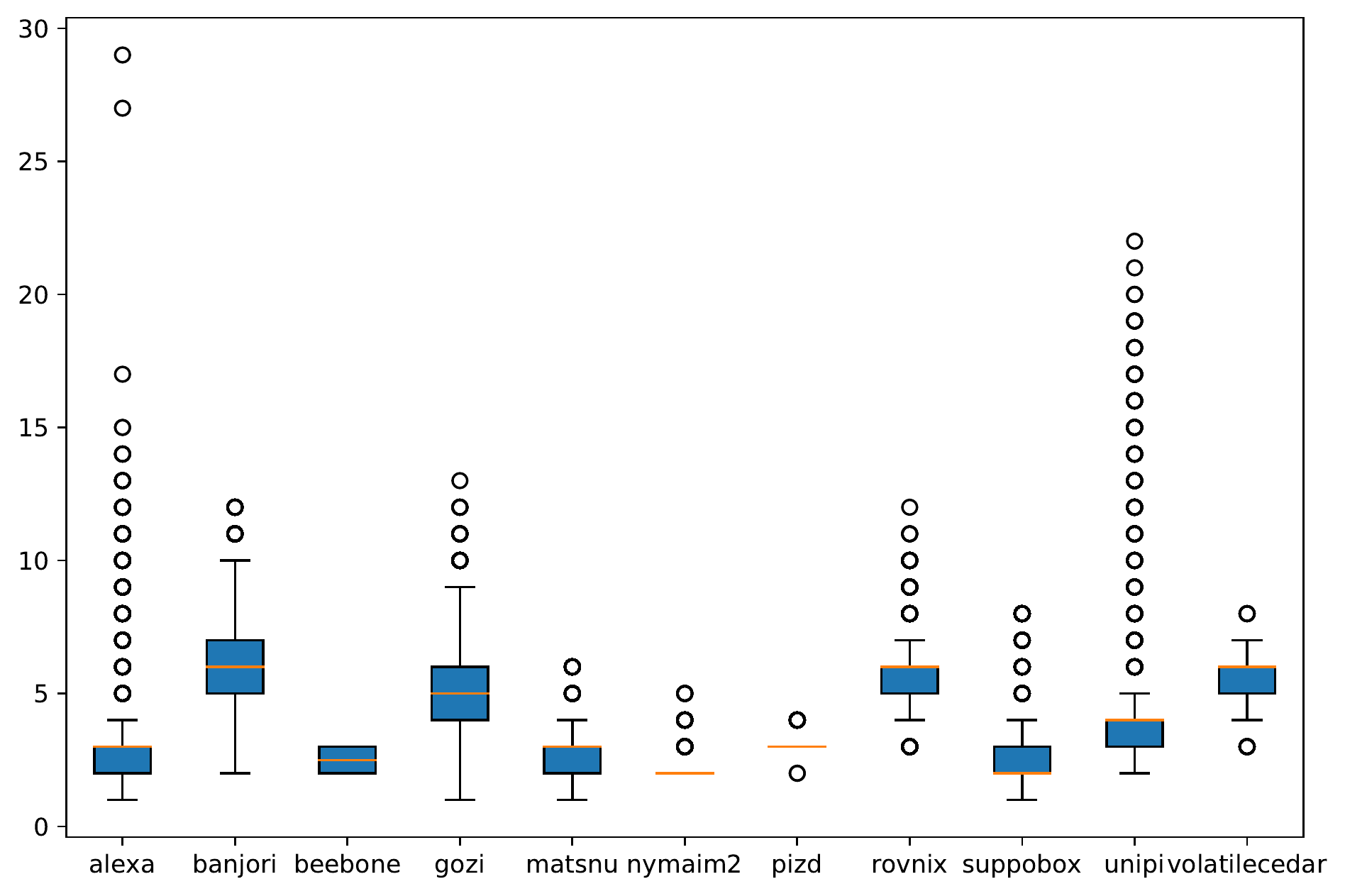}
        \caption{Average words per domain.}
    \end{subfigure}
    \caption{Textual statistics for Alexa top 100K and the DGA families studied.}
\label{fig:stats}
\end{figure*}

We conducted another set of experiments to study the statistical difference that the wordlist-based DGAs have from regular domains. Intuitively, we argue that the ``poor'' dictionary that these DGAs have would result in often repeating the same words in NXDomain queries as well as exhibiting some identifiable patterns. To this end, we analyse the textual statistical properties of the previously selected DGA families. Next, we compare them with those obtained in the case of Alexa top 1 million domains, and we depict the results in Figure \ref{fig:stats}. It should be noted that each domain in the \texttt{unipi} dataset has more or less the same length and similar amount of words with the Alexa ones.

\begin{figure}[!ht]
    \centering
    \includegraphics[width=.9\columnwidth]{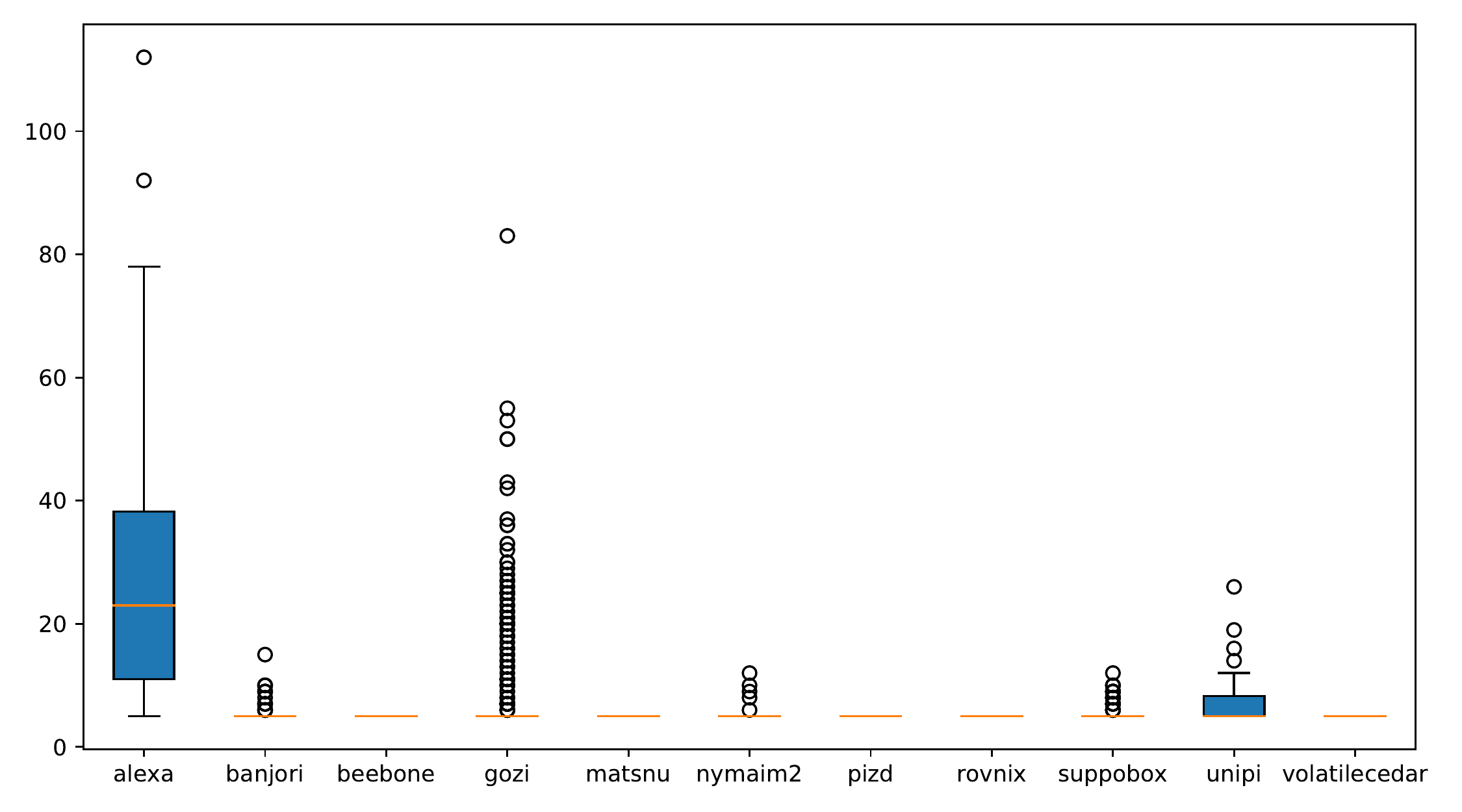}
    \caption{Number of requests needed to pass the pattern criterion threshold.}
    \label{fig:prune}
\end{figure}

Based on these statistics, we create a filter as follows. We keep a short registry of the five most recent NXDomain requests, and we check whether any of the following criteria holds:
\begin{itemize}
    \item All the requests are above ten characters.
    \item The amount of words in all requests are the same.
    \item The amount of words in all requests is above 2.
    \item The amount of the ``short'' words (less than four characters) are more than 2 in all requests.
    \item All the requests are made to the same SLD and different TLD.
\end{itemize}
The results of this process are illustrated in Figure \ref{fig:prune}. While this pattern approach introduces a bias in terms of language constraints, it is something that can be resolved by extending the dictionary of the underlying splitting algorithm. This extension may solve the issue for Latin-based dictionaries; however, this does not resolve the case of IDNs. Apparently, this filter manages to efficiently determine the lexicographical structure of the domain name. Moreover, it complements the previous filter by keeping a record of how many times a specific lexicographical structure was identified. Should these occurrences pass a threshold during a predefined epoch, the corresponding alert is raised.

As in the previous case, we performed our experiments 1,000 times. It is evident that our filter shows significant differences between AGDs, the benign (Alexa), and semi-benign (unipi). Notably, while there are some outliers for all DGAs the average of the counter is close to 5, with the highest being 5.24 from \texttt{suppobox}, while Alexa had an average of 24.19 requests, and \texttt{unipi} ranged from 5 to 25 with an average of 6.93. We believe that the above illustrates that almost all DGAs could be identified with at most six requests.

\begin{figure}
    \centering
    \includegraphics[width=.9\textwidth]{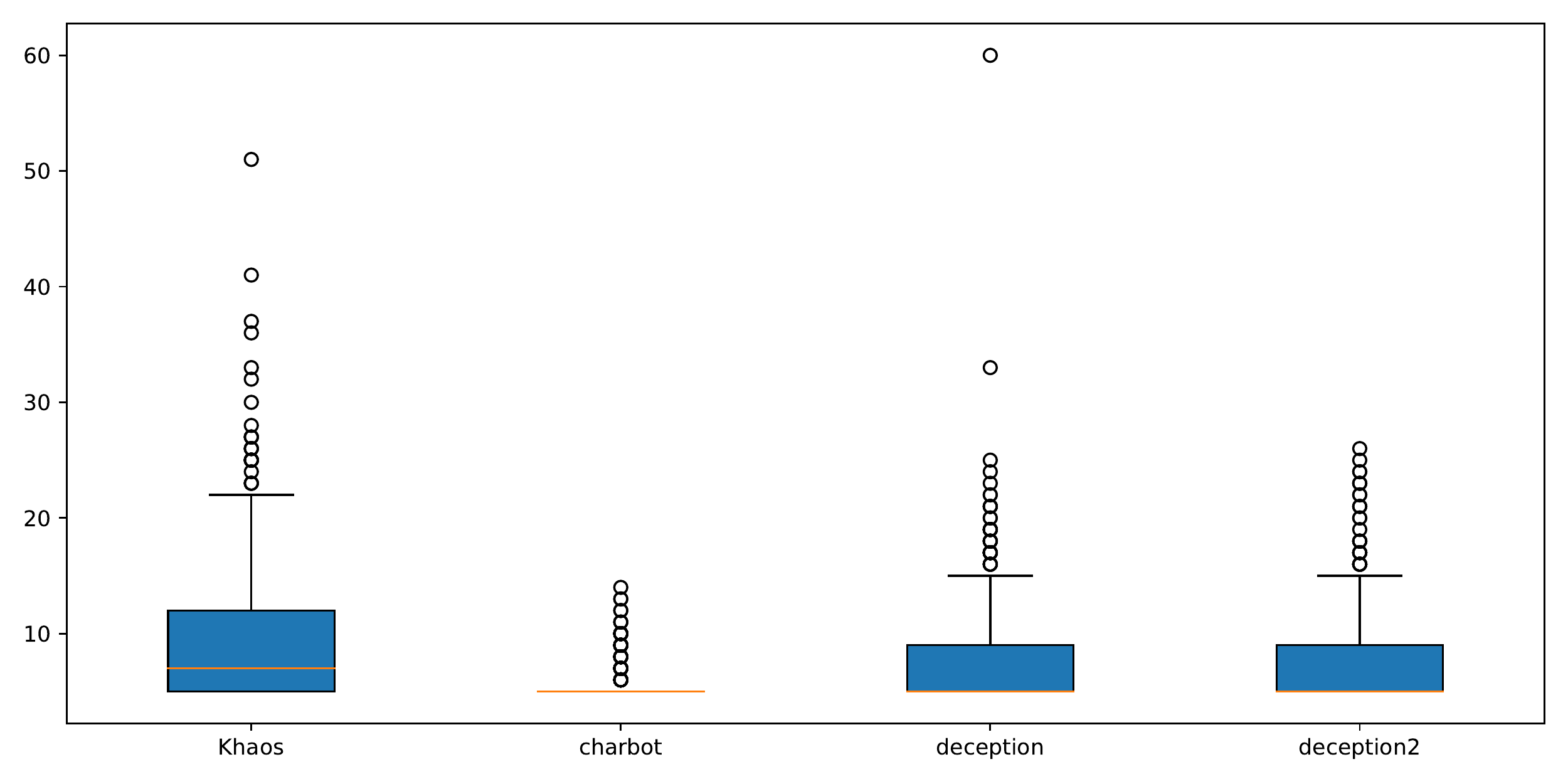}
    \caption{Number of requests needed for the artificially made DGAs.}
    \label{fig:hard2detect}
\end{figure}

Finally, we used our pattern approach on the hard to detect DGAs crafted in \cite{Spooren2019} and \cite{khaos}. Notably, All three DGAs scored significantly higher than the other DGAs. Interestingly, the average of 1,000 experiments showed that \texttt{charbot} achieved the worst performance, requiring 5.33 queries in average. In the case of \texttt{decept}, it was marginally harder than \texttt{decept2}, requiring 7.78 and 7.01 queries in average, respectively. Nonetheless, the previous are outperformed by the \texttt{khaos} DGA of Yun et al. \cite{khaos} which needed an average of 9.47 queries to be detected. The corresponding statistics are illustrated in Figure \ref{fig:hard2detect}.

\section{Discussion}
\label{sec:discussion}

In this section, we discuss the results of the experiments described in Section \ref{sec:experiments} as well as the main benefits of our approach and how it compares to the current state of the art.

In the case of the classification filter, Table \ref{tab:binary_stats} provides the classification outcomes of all the families analysed in this paper. As we previously discussed, the most difficult families to detect are the adversarial ones, with the exception of \texttt{charbot}, which is detected in almost 100\% of cases. Nevertheless, as later seen in Table 2, our approach outperforms \texttt{khaos}, \texttt{decept}, \texttt{decept2} and \texttt{charbot} detection rate reported by their corresponding original works, even if it is not crafted to detect them. Moreover, this is the first work which provides an analysis of such a rich set of wordlist-based families. The latter means that our approach is able to detect, in real-time, different DGA families created using different algorithms without modifying the classification filter, showcasing the efficiency and adaptability of our approach.

The results depicted in Figure \ref{fig:strikes} show that all the DGA families evaluated can be detected by our word-based filter with only a few NXDomain queries (each domain resulting in a set of processed words), except \texttt{nymaim2}. This means that the AGDs generated by them tend to repeat words with statistical significance, as detected by our word-based filter. In the simplest case (i.e. with a threshold of 3 words, cf Figure \ref{fig:strikes}a), we can detect AGDs with less than 30 NXDomain queries in almost all cases. For instance, the malware \texttt{gozi} uses 3 times the same word after generating 28 domains (see Figure \ref{fig:strikes}a). In addition, the growth pace exhibited in Figure \ref{fig:strikes}, in terms of NXDomain queries needed to reach from 3 to 7 strikes, exhibits the same growth pace as the probabilities defined in Section \ref{sec:methodology}. The latter implies that, proportionally, the number of strikes grows faster than the number NXDomains analysed.

In the case of \texttt{nymaim2}, the results show that we need a high amount of NXDomains to find coincidences (see Figure \ref{fig:nymaim2detail}). This occurs because \texttt{nymaim2} uses a predefined structure to create domains in which two words, selected from two separate dictionaries with 2450 and 4387 words, respectively, are appended to a TLD (i.e. the number of possible TLDs is 74). Therefore, the amount of possible combinations hinders its detection. Nevertheless, this variability is only in the dictionary, and since the structure remains the same, it is captured by our pattern filter.  It is worth noting that \texttt{beebone} needs a constant number of queries to be detected by our word filter as the words that act as prefix/suffix are constant (cf Table \ref{tbl:sample}) in all queries. Therefore, they both trigger the alert in as many queries as the threshold is.

It is clear from Figure \ref{fig:stats} that benign domain names most likely consist of at most three words containing less than ten letters in total. Therefore, NXDomain requests that do not meet these criteria can be considered as `suspicious' by our pattern-based filter. Our claim is verified by the results of our pattern filter in Alexa and \texttt{unipi} (cf. Figure \ref{fig:prune}), since the number of domains needed to pass the threshold is far higher than those required by the rest of DGAs. Note that the pattern-based filter can fully capture \texttt{nymaim2} behaviour (as well as the rest of families, with few exceptions) so that we can raise an alert faster than our word-based filter in such cases.

Of specific interest is the \texttt{unipi} dataset. As already mentioned, the dataset consists of NXDomain requests from our institution's network. The extracted network traffic contained 253 NXDomain requests that originated from DGAs. Therefore, at least 7\% of the dataset is malicious. Nonetheless, this was quickly detected from our filters and is depicted in the significant difference between the Alexa and \texttt{unipi} dataset in the experiments reported in Section \ref{sec:experiments}.

Finally, it is worth noticing that the bulk of hash and arithmetic-based DGAs, such as \texttt{Dyre}, \texttt{Gameover}, \texttt{Gspy} and \texttt{Omexo}, produce domains that are long (more than 15 characters) and as they are hex-encoded, they hardly create words. Therefore, all such AGDs fail our structure criterion. Moreover, AGDs generated by other DGA families, including \texttt{DNS Changer}, \texttt{DiamondFox}, \texttt{DirCrypt} and \texttt{EKforward} also fail our structure criterion, as they might generate shorter domains, but the produced domains do not have meaningful words.

\begin{sidewaystable}[!ht]
\centering
\tiny
\rowcolors{2}{gray!25}{white}
\begin{tabular}{lccccccccccc}
\toprule
DGA & Koh & Maldon & Berman  & Curtin  & Xu  & Yang   &  Yun & Spooren & Peck & \multicolumn{2}{c}{\textbf{Our Method}}  \\
& et al.\cite{koh2018inline} & et al.\cite{ALMASHHADANI2020101787} & et al.\cite{berman2019dga} & et al.\cite{curtin2018detecting}& et al.\cite{XU201977} & et al.\cite{yang2019detecting}& et al. \cite{khaos}& et al. \cite{Spooren2019} & et al. \cite{8756038}  &  $q<t$ & $q>=t$ \\
\midrule
beebone       &                           &                           & 100                  & 74.90                &                           &                           &                           &                             &                           & 99.65                          & 100                             \\
banjori       &                           & \multicolumn{1}{r}{90.57} & 100                  & 80.80                & \multicolumn{1}{c}{99.78} &                           &                           &                             &                           & 99.74                          & 100                             \\
charbot       &                           &                           & \multicolumn{1}{l}{} & \multicolumn{1}{l}{} &                           &                           &                           &                             & \multicolumn{1}{c}{98.89} & 98.93                          & 100                             \\
khaos         &                           &                           & \multicolumn{1}{l}{} & \multicolumn{1}{l}{} &                           &                           & \multicolumn{1}{c}{80.30} &                             &                           & 86.73                          & 100                             \\
decept        &                           &                           & \multicolumn{1}{l}{} & \multicolumn{1}{l}{} &                           &                           &                           & \multicolumn{1}{c}{85.72}   &                           & 89.31                          & 100                             \\
decept2       &                           &                           & \multicolumn{1}{l}{} & \multicolumn{1}{l}{} &                           &                           &                           & \multicolumn{1}{c}{79.45}   &                           & 85.70                          & 100                             \\
gozi          &                           & \multicolumn{1}{r}{78.32} & 0                    & 77.30                & \multicolumn{1}{c}{97.97} &                           &                           &                             &                           & 92.87                          & 100                             \\
matsnu        & \multicolumn{1}{c}{94.40} &                           & 0                    & 89.10                & \multicolumn{1}{c}{96.78} & \multicolumn{1}{c}{93.01} &                           &                             &                           & 89.94                          & 100                             \\
nymaim2       &                           &                           & \multicolumn{1}{l}{} & \multicolumn{1}{l}{} &                           &                           &                           &                             &                           & 89.69                          & 100                             \\
pizd          & \multicolumn{1}{c}{94.24} &                           & 91.96                & \multicolumn{1}{l}{} &                           &                           &                           &                             &                           & 93.61                          & 100                             \\
rovnix        & \multicolumn{1}{c}{95.59} & \multicolumn{1}{r}{99.63} & \multicolumn{1}{l}{} & 80.50                &                           &                           &                           &                             &                           & 98.68                          & 100                             \\
suppobox      & \multicolumn{1}{c}{95.69} &                           & 87.82                & 56.80                & \multicolumn{1}{c}{98.19} & \multicolumn{1}{c}{86.23} &                           &                             &                           & 88.49                          & 100                             \\
volatilecedar &                           &
& 100                  & 95.80                &                           &                           &                           &                             &                           & 98.52                          & 100                            \\

\bottomrule
\end{tabular}
\caption{Comparison of the detection performance (in percentage) between state-of-the-art methods. In each case, we report the best values.}
\label{tab:comparison}
\end{sidewaystable}

To showcase the efficacy of our method, we compared it with the most relevant state-of-the-art. Note that, despite comparing the same DGA families, the datasets reported in the related works may have different samples and/or they are not reproducible and thus, a thorough comparison with such works is not feasible. The use of different datasets and the lack of updated benchmarks is a well-known issue in DGA research, due to the continuous introduction of new DGA families. Nevertheless, in Table \ref{tab:comparison}, we reported the best outcomes for each related work, according to either Accuracy or F1-Score metric. It can be observed that none of the methods compared includes the whole list of word-based DGAs analysed in this article, as stated in the Motivation section. We can observe that most of the methods succeed to provide a remarkable detection rate in at least one of the families. Moreover, in many cases, the size of the samples evaluated is extremely small (i.e. which evinces the difficulty to obtain a quality database such as the one used in this article).

After a deeper analysis of the outcomes, we observed a common trend in most of them. More concretely, when a method can accurately detect a DGA family, it fails to detect others, due to the particular characteristics of each DGA, as seen in \cite{berman2019dga,curtin2018detecting,yang2019detecting,ALMASHHADANI2020101787}. Note that there are cases where the method is not able to capture any instance of a DGA (i.e. the reported detection rate is 0). In addition, some of the families are only explored in this paper (i.e. \texttt{nymaim2}) as well as \texttt{decept}, \texttt{decept2}, \texttt{charbot} \texttt{khaos}, which are only analysed by their creators \cite{Spooren2019,khaos,8756038} by using state-of-the-art methods. More precisely, in the case of \texttt{decept} and \texttt{decept2} the detection rate is below 85\% using LTSM, and in the case of \texttt{khaos}, the authors reported that they detected the AGDs in the 80.30\% of cases. Our experiments show that we can efficiently detect \texttt{charbot}, \texttt{decept}, \texttt{decept2}, and \texttt{khaos}, with higher accuracy by just using the classification filter. Moreover, as described in Section 3, the classification filter is analysing in real-time the characteristics and features of each domain name. In the case of reaching the bucket thresholds, we assume that our method is fully capturing the DGA with no exception, due to the birthday paradox and the outcomes of our experiments in Section 4. Therefore, as described in Table \ref{tab:comparison}, the detection rate of our method is equal to the classification filter outcome if $q<t$, where $q$ is the number of queries required to reach the bucket threshold (this varies according to each family, as seen in Figure \ref{fig:strikes}), and $t$ is the threshold value. In the case of $q>=t$, we report a 100\% detection rate.

Finally, it is worth mentioning that when some DGAs change their seed, the accuracy of their detection drops significantly, as stated by Berman \cite{berman2019dga} in the case of \texttt{pizd} and \texttt{suppobox}. However, the latter behaviour does not affect our method, since it only implies a restart of the word counter. In addition, we did not include the works presented in \cite{pereira2018dictionary} and \cite{lison2017automatic} due to their database size, since they train their methods with a high amount of domains, in some cases orders of magnitude higher than the queries needed by our method to detect them. For instance, in \cite{pereira2018dictionary}, authors use 8-day real traffic data, resulting in thousands of domains. In the case of \cite{lison2017automatic}, they use tens of thousand queries generated by \texttt{banjori}, \texttt{gozi}, \texttt{nymaim2}, \texttt{rovnix}, \texttt{suppobox}, \texttt{matsnu} (i.e. with low detection accuracy in the case of \texttt{matsnu}, which is below 0.16), and hundreds of queries in the case of \texttt{beebone} and \texttt{volatilecedar}, to train their system using a 10-fold scheme. Another recent work, proposed by Yang et al. in \cite{yang2019detecting}, uses machine learning and semantic analysis to detect two DGA families, namely \texttt{suppobox} and \texttt{matsnu}. Using different dataset configurations, they report accuracies between 83.63\% and 86.23\% for \texttt{suppobox} and between 88\% and 93.01\% in the case of \texttt{matsnu}. Although they also consider an additional set of unclassified word-based AGDs, their average detection accuracy is 80.58\%, which is significantly lower than our proposed work. Finally, in work presented by Spooren et al. \cite{Spooren2019}, the authors achieve different accuracies depending on the classifier and the parameters. In general, the random forest classifier is rendered useless for \texttt{decept} and \texttt{decept2} with accuracies below 60\%. In the case of LTSM, the classification accuracy for such families is also lower than ours. Nevertheless, as seen in our experiments, the performance key of such classifier is the proper use of a rich set of features which represents relevant statistical patterns.

In summary, in addition to absolute accuracy in DGA detection, our method enables a set of benefits, compared to other well-known literature methods based on neural networks or other feature-based classification mechanisms \cite{koh2018inline,18,pereira2018dictionary,lison2017automatic,yang2019detecting,Spooren2019,ALMASHHADANI2020101787}. More concretely, our method can be deployed instantly and is parallel by design. In addition, the threshold-based patterns do not require training, contrary to the aforementioned approaches, enabling the adaptable discovery and detection of novel DGA families. Note that, since they require training, Neural Networks are sensitive to dictionary changes, hence providing less robust outcomes for real-time DGA analysis than our approach. As a further enhancement, our pattern-based setting can filter out almost instantly DGAs, which means that only a minimal subset needs to be further analysed. Moreover, our word-based filter only needs a small number of NXDomain queries to achieve DGA detection. This also enables personalised policies, where benign domains that fall out of the threshold can be whitelisted since their amount would be relatively small. We also argue that the fast detection of \texttt{decept}, \texttt{decept2} and \texttt{khaos} signify another advantage of our methodology. More precisely, while these DGAs are crafted to exploit many character features, they fail to be meaningful enough and end up being detected by our classification and pattern filters. Finally, paired with recent advances in human typewriting error detection and application layers preventing such behaviour, the robustness of our method can be highly enhanced.

It is worth noting that the proposed method is directly affected by the language of the dictionaries and their size. Large dictionaries, as seen in the case of \texttt{nymaim2}, imply a longer detection threshold. On the other hand, the case of using other languages for the dictionaries implies another issue for the method. Nevertheless, the latter is the same for all methods targeting these DGAs, as they all depend on the knowledge of the used language to split the words accordingly and perform their analysis. In this regard, one strategy to circumvent our approach could entail the use of large dictionaries based on, e.g. Alexa words. In this setup, our system would require from larger thresholds, and most of the times the resolved domains would not generate an NXDomain response. Nevertheless, this implies a set of risks for an attacker, since finding the C\&C server would require a higher number of queries, reducing the practicality and efficiency of the attack and increasing the detection risk due to the suspicious amount of queries.

Another, more generic, approach to bypass the NXDomain filters is the use of covert and encrypted communications. Nevertheless, this approach has been already studied in \cite{PATSAKIS2020101614}, where authors described that NXDomain responses could be properly captured and patterns could be learned from such encrypted response data to detect AGDs accurately. Further approaches to bypass the methodology could entail the use of different combinations of n-grams following, e.g. English statistical patterns, so that a large subset of features would be similar for both benign and malicious domain names. The latter would also hinder the probability-based filters of our approach, yet new features could be designed to train the classification filter in such case.

\section{Open challenges and final remarks}
There are several challenges to overcome in the DGA-based research field, and these are mainly related to the quality and reproducibility of the experiments. In contrast to other research fields, such as computer vision or medical diagnosis, which rely on standardised and well-known benchmarks, the DGA-based benchmarks need to be continuously updated due to the appearance of novel DGA families. The latter implies that for the proper assessment of a methodology in real scenarios, the system should be tested against the most recent DGA families. Nevertheless, the creation of a rich and balanced dataset containing all possible families is a hard task \cite{ZAGO2020101719}, since some of the families have not been reverse engineered yet, or because their inherent implementation only creates a small subset of combinations. Therefore, we believe that the creation of upgradable benchmark versions is a requirement for this research field.

Another issue that arises in the DGA literature is the lack of standardised metrics to enable robust comparisons. In this regard, after analysing the works described in Section 2, we observed that several authors used ad-hoc benchmarks or feeds that are no longer available (note that authors report the use of well-known sources, yet the feeds change daily or weekly) \cite{yu2017inline,mac2017dga,3,chen2018towards}, specific metrics which hinder the comparison with other works (e.g. not using precision, recall, $F_1$-score, or accuracy) \cite{yu2017inline,curtin2018detecting,18,khehra2018botscoop}, and finally, reporting the binary classification outcomes in an aggregated way  \cite{zago2019scalable,khehra2018botscoop,ZAGO2020101719,ALAEIYAN2020,ALMASHHADANI2020101787,9072447,18,lison2017automatic,3,chen2018towards,ANAND20201129}. The latter is particularly relevant to evaluate the quality of a method against specific sets of families since approaches might obtain highly accurate results for some families, while they are unable to detect other families. Moreover, an unbalanced amount of AGD samples per family can also bias the aggregated outcome.

\begin{figure}[t]
    \centering
    \includegraphics[width=0.8\textwidth]{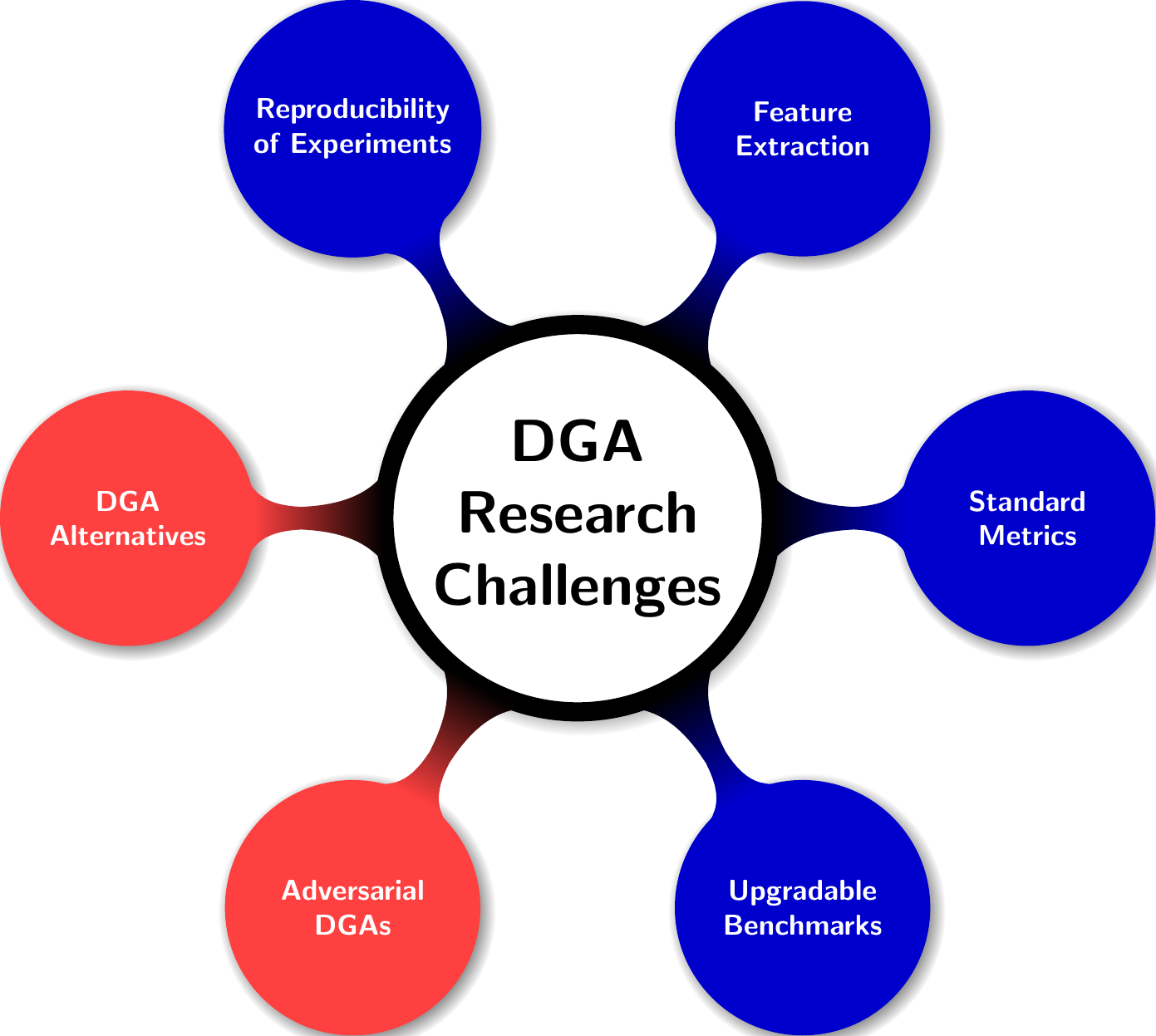}
    \caption{Mindmap representation of the main challenges of the DGA research. }
    \label{fig:tikz_challenges}
\end{figure}

From the adversary's point of view, the main research lines, as they also have to be investigated, there are two main streams: adversarial DGAs and alternatives. In the former case, the adversary tries to exploit the knowledge about the machine learning algorithm which is used to detect these domains. This way, the adversary may try to, e.g. create AGDs with specific n-gram properties. Recently, several researchers have followed this train of thought introducing several such DGA, some of which are analysed in this work. Therefore, this showcases the need for exploring adversarial training and other such methods in this field, as well as extracting relevant features to identify them. Finally, one has to consider that DGAs might be a mechanism widely used by malware, however, they may soon switch to alternatives to exploit, e.g.  decentralised platforms, as highlighted in \cite{Patsakis2019}. Figure \ref{fig:tikz_challenges} summarises the main challenges of the DGA research field.

In this work, we analyse the current state of the art in Domain Generation Algorithms, a family of algorithms that use pseudo-random generators to create a set of AGDs, used as rendezvous points for their C\&C servers. More concretely, we focus on DGA families which use wordlists to generate such domains. Therefore, to provide efficient and accurate detection of wordlist-based DGAs, we propose a probabilistic method inspired by the ``birthday paradox'' and the structure that these generators have. In this regard, our method exhibits a series of benefits compared with other state-of-the-art methods, since it can be instantly deployed, leverages real-time classification, and several of its modules do not require training, as they can filter the bulk of domain names in terms of their pattern construction, enabling efficient and adaptable DGA detection. Moreover, extensive experiments using state-of-the-art benchmarks show that we need between 3 and 27 NXDomain queries (with strike threshold set to 3) to detect DGA malware with high confidence using our word-based filter. Future work will focus on analysing the statistical properties of benign domains (including IDN), especially in the case of Alexa, to enhance their classification using different wordlist-based probabilistic word splitters.

\section*{Acknowledgments}
This work was supported by the European Commission under the Horizon 2020 Programme (H2020), as part of the project YAKSHA (Grant Agreement no. 780498) and CyberSec4Europe (\url{https://www.cybersec4europe.eu}) (Grant Agreement no. 830929), \textit{LOCARD} (\url{https://locard.eu}) (Grant Agreement no. 832735).

The content of this article does not reflect the official opinion of the European Union. Responsibility for the information and views expressed therein lies entirely with the authors.

\section*{References}
\bibliographystyle{elsarticle-num}
\bibliography{biblio}

\end{document}